\documentclass[10pt, floatfix, twocolumn, aps, prr, longbibliography, groupedaddress, superscriptaddress]{revtex4-2}

\usepackage{xcolor}
\usepackage{amsmath}
\usepackage{amssymb}
\usepackage{graphicx}
\usepackage[english]{babel}
\usepackage{textcomp}
\usepackage{braket}
\usepackage{mathtools}
\usepackage{dsfont}
\usepackage{comment}
\usepackage{diagbox}
\usepackage{wrapfig}
\usepackage{hyperref}
\usepackage{lineno}

\newcommand{\fref}[1]{Fig.~\ref{#1}}

\newcommand{\dgr}{^{\dagger}}

\newcommand{\enum}[1]{\mathit{e}^{#1}}

\begin{document}
\title{Encoding a topological gauge theory on a ring-shaped Raman-coupled Bose gas}

\author{Claudio Iacovelli}
\email{claudio.iacovelli@icfo.eu}
\affiliation{ICFO - Institut de Ciencies Fotoniques, The Barcelona Institute of Science and Technology,
Av. Carl Friedrich Gauss 3, 08860 Castelldefels (Barcelona), Spain}
\affiliation{Departament de F\'isica, Universitat Aut\'onoma de Barcelona, E-08193 Bellaterra, Spain}
\author{Josep Cabedo}
\affiliation{Departament de F\'isica, Universitat Aut\'onoma de Barcelona, E-08193 Bellaterra, Spain}

\author{Leticia Tarruell}
\affiliation{ICFO - Institut de Ciencies Fotoniques, The Barcelona Institute of Science and Technology,
Av. Carl Friedrich Gauss 3, 08860 Castelldefels (Barcelona), Spain}
\affiliation{ICREA, Pg. Llu\'is Companys 23, 08010 Barcelona, Spain}
\author{Alessio Celi}
\email{alessio.celi@uab.cat}
\affiliation{Departament de F\'isica, Universitat Aut\'onoma de Barcelona, E-08193 Bellaterra, Spain}

\begin{abstract}
Topological gauge theories constitute a framework for understanding strongly correlated quantum matter in terms of weakly interacting composite degrees of freedom. Their topological properties become evident when these theories are realized on a space of non-trivial topology. Here, we propose a scheme to realize a one-dimensional topological gauge theory, the so-called chiral BF theory, on a ring geometry. We obtain such a theory by dimensionally reducing Chern-Simons theory on a disk to the chiral BF theory defined on the ring. Then, we encode the theory into a Hamiltonian with a coupling between angular momentum and density, and we propose and numerically benchmark its realization in an optically-dressed Bose gas confined in a ring-shaped trap. There, the topological properties of the underlying theory manifest themselves through a magnetic flux variable that is density-dependent. We quantify such density-dependent magnetic flux in terms of the ground-state angular momentum and the chiral properties of the system through a Bogoliubov analysis. Our proposal enables the observation of topological features of the chiral BF theory that become manifest due to the non-trivial topology of the ring geometry. 
\end{abstract}

\maketitle

\section{Introduction}\label{Introduction}
Topological gauge theories (TGTs) are a special class of gauge theories with no ``photons'', that is, at the classical level the gauge field is not propagating in the vacuum \cite{TFT1,TFT2}. The classical action is a {\it topological invariant} and  assigns a set of local conservation laws that determines gauge field configurations in terms of the matter ones, up to global solutions, i.e, related to the topology of spacetime. 
In this paper, we propose a simple ultracold atom \cite{Lewenstein2012} setup to investigate experimentally the role of spacetime topology in a textbook example of topological gauge theory, namely the chiral BF theory \cite{Rabello1995,Rabello1996,Aglietti1996}.

In high energy physics, topological contributions to gauge theory actions are included to provide a non-perturbative completion of the Standard Model (such as the so-called $\theta$-term) \cite{Peskin1995,tHooft, Cottingham_Greenwood_2007}, or to describe physics beyond the Standard model \cite{axion2007}. In condensed matter, topological gauge theories provide low-energy descriptions of strongly-correlated quantum materials through effective weakly interacting models  \cite{Ezawa2000, Fradkin2013, Sachdev2023}.
A prominent example of this are fractional quantum Hall systems. There, fractionalized Hall conductance is explained through quasi-particle excitations which can be described as electrons in a plane with an emerging magnetic field attached to them, the famous flux attachment \cite{Wilczek1982,Jain1989}.
In the continuum limit, the flux attachment condition implies that the magnetic field is proportional to the charge density, and can be interpreted as the conservation law of Chern-Simons theory \cite{ZhangKivelson1989,JackiwPi,Iengo1991}, the topological gauge theory for the emergent gauge field generating the magnetic flux.   
Instead of Chern-Simons theory, here we consider the chiral BF theory, a one-dimensional (1D) gauge theory that can be thought of as a dimensional reduction of Chern-Simons on its edge \cite{Rabello1995,Rabello1996,Aglietti1996,Jackiw1997,griguolo1998}. The name “chiral BF theory” has the following origin. The term “BF” originates from the topological coupling between a bosonic field and the gauge-field strength, which in $1+1$ dimensions takes the form of a scalar field $\mathcal{B}$ multiplied by $\varepsilon^{\mu\nu}F_{\mu\nu}$. The term “chiral” reflects the fact that the kinetic term of $\mathcal{B}$ is equivalent to that of a chiral boson.

As first proposed by \cite{Rabello1995} and clarified by \cite{Kundu1999}, the chiral BF model provides a gauge theory description of a one dimensional anyon gas and, as shown by \cite{Bonkhoff2020}, it corresponds to the continuum formulation of the anyon Hubbard model \cite{Yajiang2009,Keilmann2011}.
Recently, the chiral BF theory on a line was realized in \cite{Frolian2022} with a two-component Raman-coupled Bose-Einstein condensate (BEC) by experimentally demonstrating the existence of chiral solitons and the asymmetric expansion of the gas. Moreover, the anyon Hubbard model was realized in \cite{Greiner2024} for two particles and their asymmetric transport was observed. Anyonic correlations have also been observed through the momentum distribution of an impurity travelling in a Tonks-Girardeu gas \cite{Dhar2025}. These experiments observed the chirality routed in 1D anyons and in their field theory description, the chiral BF theory. However, these experiments could not access topological invariants characterizing it because the topology of the space was trivial.
\begin{figure*}[t!]
\centering
\includegraphics[width=1.0\linewidth]{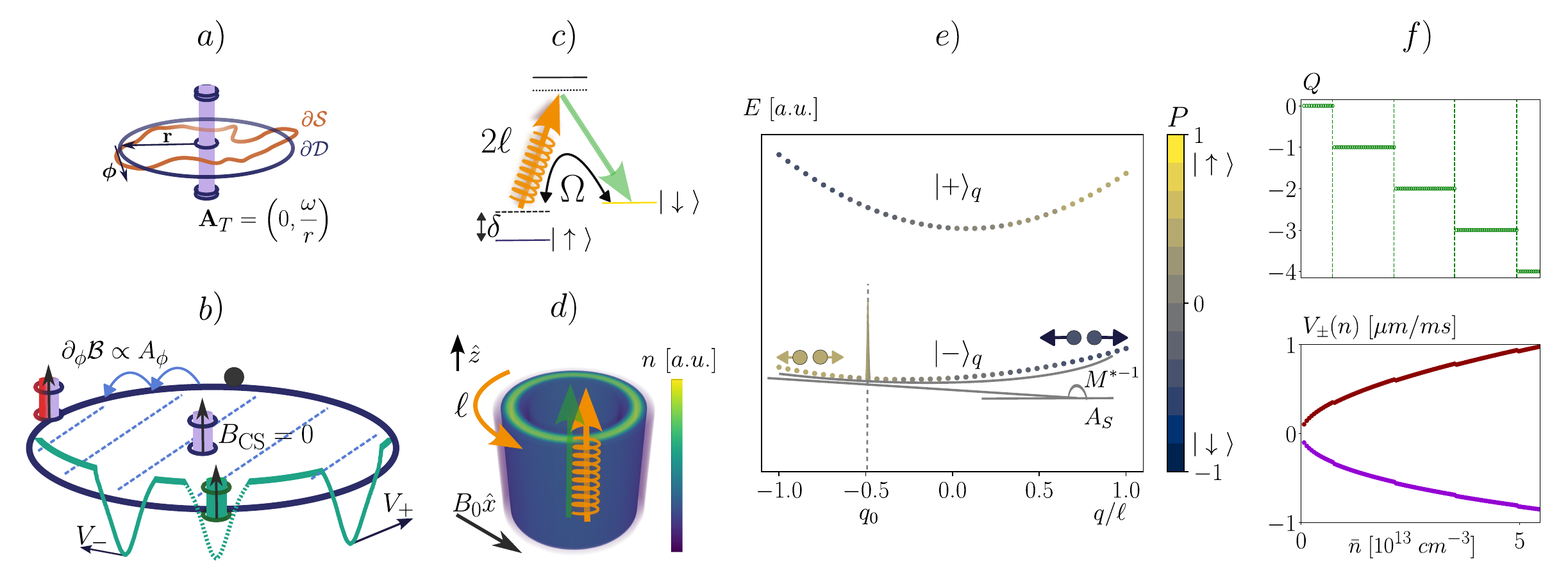}
\caption{Chiral BF theory on a ring: derivation from Chern-Simons theory, experimental scheme and signatures. a) A winding number $\omega\in\mathbb{Z}$ is equivalent to placing an infinitesimal solenoid in the origin; outside the magnetic field is everywhere zero but there is a flux $2\pi\omega$. b) Lagrangian on a disk and its encoded Hamiltonian. Chern-Simons theory is represented as a purple solenoid in the bulk and the BF term as the red-shaded half of the solenoid outside the boundary. It restores gauge invariance by completing the solenoid cut by the edge. Integrating Chern-Simons theory in the bulk gives rise to a chiral boson term on the boundary (skipping orbits). Coupling the theory to a scalar field (black sphere) yields a Hamiltonian with chiral interactions, i.e., the angular momentum coupled to a density-dependent magnetic flux (green solenoid),
which manifest in different sound velocities for right and left movers ($V_{\pm}$). c)-d) Experimental implementation of the encoded Hamiltonian. c) A Laguerre-Gauss (orange) and a Gaussian (green) beams produce a Raman transition between internal states $\ket{\uparrow}$ and $\ket{\downarrow}$ of a Bose gas with single-particle detuning $\delta$ and a two-photon Rabi coupling $\Omega$. The laser scheme simultaneously produces a scalar ring potential.  d) Density profile of the gas with an angular momentum transfer $\ell$, together with the beams' direction of propagation and the quantization axis given by a bias magnetic field $\mathbf{B} = B_0 \hat{x}$. e) Chiral BF theory as effective Hamiltonian. Dispersion bands of the single-particle Hamiltonian as a function of the quasi-angular momentum $q$. The colormap shows the spin polarization $P$ of the eigenstates. A Taylor expansion of the lower band around a center momentum $q_0$ yields a static gauge $A_S$ (the slope of the tangent) and an effective mass $M^*$ (the curvature). Chiral interactions arise from imbalanced intrastate interactions of $\ket{\uparrow}$ and $\ket{\downarrow}$, producing a density-dependent displacement of the lower band, which is depicted by particles repelling with different strengths at different momenta of the band (sketches with spheres and arrows). f) Experimental signatures of the theory. Change of the angular momentum $Q$ (upper panel) and the chiral sound velocities $V_{\pm}$ (lower panel) as a function of the density-dependent magnetic flux variable (see Section \ref{sec:numerics}).
}
\label{Fig_1}
\end{figure*}

In this work, we address the lack of topological effects by considering the chiral BF theory on a ring. The ring hosts topological configurations of the gauge field, i.e., extra, zero-magnetic field configurations with an integer-valued magnetic flux which exist also in the absence of matter. We show the emergence of the chiral BF theory from Chern-Simons theory on a disk by making two consecutive observations. First, the boundary term required to restore gauge invariance on a finite domain \cite{Wen1992,Dunne1999} is the BF term with an extra degree of freedom $\mathcal{B}$. Then, the integration of the Chern-Simons disk along the radius returns the chiral boson action for $\mathcal{B}$, thus recovering the full chiral BF theory in the absence of matter. After including the coupling to matter, we reformulate the theory into a Hamiltonian by means of the Fadeev-Jackiw procedure \cite{FaddeevJackiw1988,Jackiw1993,Craig2022}, which removes the redundant gauge degrees of freedom by encoding them into matter. This encoded Hamiltonian features chiral interactions in the form of a coupling between angular momentum and a magnetic flux proportional to the matter density. To engineer such a coupling, we propose to use a ring-shaped Bose Einstein condensate that is Raman-coupled  \textit{via} a Laguerre-Gauss beam \cite{Edmonds2013,Ohberg2016} and employ the same description as in \cite{Craig2022} for the Raman-coupled system.
Ultracold gases in a ring trap have been extensively studied for inspecting the properties of persistent currents in quantum gases, as well as for atom interferometry and atomtronics \cite{Ryu2007, Moulder2012, Moulder2013, Eckel2014,Pandey2019, Guo2020, Ryu2020,Amico2021,Amico2022}. Laguerre-Gauss beams have been also proposed to selectively excite and detect edge and collective modes in fractional Chern-insulators engineered with ultracold atoms \cite{Goldman2012, Goldman2024}. In turn, Raman coupling in simply connected geometries has become a key tool to engineer synthetic vector potentials \cite{Goldman2014}, spin-orbit coupling \cite{Spielman2011,HoZhang2011, Galitski2013} and tunable interactions \cite{Williams2012}. Recent experiments have also realized spin-orbital-angular-momentum coupling in ultracold gases, demonstrating azimuthal gauge potentials, controlled vortex creation, and ground-state phase transitions \cite{Chen20181,Chen20182,Zhang2019}. Here, we propose to exploit Raman coupling on a ring geometry to achieve emergent gauge fields that are azimuthal. Consequently, their circulation along the ring is a magnetic flux variable. We show that the chirality of interactions achievable in Raman-coupled Bose gases leads to a magnetic flux that is density-dependent, resulting in a system that realizes the chiral BF theory on a ring. By demonstrating a feasible scheme to realize density-controlled magnetic fluxes, our work bridges the gap between realizations of one- and two- dimensional topological gauge theories.

In order to formulate the chiral BF theory in a ring in a form amenable to quantum simulation, discuss its emergence in Raman-coupled Bose gases, and propose realistic protocols for observing the density-controlled magnetic fluxes associated to it, this work is organized as follows.  
In Section II, we review the connection between the chiral BF model and Chern-Simons theory, and deduce the former as a dimensional reduction of the latter. In Section III, we apply the Faddeev-Jackiw formalism to the chiral BF Lagrangian and obtain a Hamiltonian in second-quantized and encoded form. In Section IV, we detail the implementation of this Hamiltonian in a Bose gas with imbalanced interactions coupled by Laguerre-Gauss Raman laser beams. In Section V, we investigate the validity of the mapping to the chiral BF model for realistic experimental parameters by means of numerical simulations. Finally, in Section VI we present our conclusions and discuss the perspectives opened by this~work.
\section{Chiral BF theory on a ring}\label{ChiralBF_theory}

In this section, we will derive the topological modes of the gauge field by solving the flux attachment condition in a non trivial space. The space with non trivial topology is the plane without the origin, and the topological modes are produced by an infinitesimal solenoid which is placed at the origin. Then, we will argue that the Chern-Simons Lagrangian on a disk requires an additional boundary degree of freedom to be gauge-invariant, and show how this extended theory is dimensionally reduced to obtain the chiral BF Lagrangian on a ring.   

To derive the topological modes, we first introduce the flux attachment condition, i.e., the Chern Simons local conservation law in the absence of matter \cite{JackiwPi,Iengo1991}
\begin{equation}
\varepsilon^{ij}\partial_iA_j=\mathrm{B_{CS}}   =0\label{eq:flux_attachment_vacuum}.
\end{equation}
Here, $\varepsilon^{ij}$ is the Levi-Civita symbol, $A_i$ is the (two-dimensional) Chern-Simons gauge field $i$-th spatial component with $i=1,2$, and $\mathrm{B_{CS}}$ the Chern-Simons magnetic field. Then, a gauge field configuration $\mathbf{A}^T$ is topological when it solves Eq.\eqref{eq:flux_attachment_vacuum} but exhibits a non-zero magnetic flux $\Phi$  which is independent of the shape of the surface $\mathcal{S}$ that is used to compute it 
\cite{geometrybook}. As sketched in \fref{Fig_1} a), this condition can be fulfilled $\forall \mathbf{x}\neq \mathbf{0}$, with $\mathbf{x}$ denoting a point in the infinite plane, by placing an infinitesimal solenoid at the origin, and, if $\mathcal{S}$ encloses the origin, it follows from Stokes' theorem that
\begin{equation}
\Phi=\oint_{\partial\mathcal{S}}\mathbf{A}^T\mathrm{d}\mathbf{s}=2\pi\omega,\quad\omega\in\mathbb{Z}.\label{eq:flux}
\end{equation}
Noticeably, $\omega$ is the Chern number of the gauge field. We set $\mathcal{S}$ to a disk $\mathcal{D}$ of radius $r_0$ and look for a way to fulfill Eq.\eqref{eq:flux} on the ring  $\partial\mathcal{D}$. We choose $A^T_\phi$ that is constant along $\partial\mathcal{D}$ and evaluate the circulation as $2\pi r_0 A_\phi^T$, finding
\begin{equation}   A^T_\phi(r_0)=\frac{\omega}{r_0}.\label{eq:TAP}
\end{equation}
We prolong this solution to the whole disk, excluding the origin, by letting $r_0\to r$.  If this configuration satisfies Eq.\eqref{eq:flux_attachment_vacuum}, we infer that $A^T_r(r,\phi)=0$, as expected. In fact, there are no windings along the radial coordinate since it is not a compact dimension.

This solution also results from the flux attachment in the presence of a point matter source at the origin \cite{Tong2016}. 
With this construction, we observe that the non trivial topology of the space, i.e. $\mathbb{R}^2\backslash\{\mathbf{0}\}$, implies an additional flux variable but no corresponding magnetic field. We will call $A^T_\phi$ the solenoid solution, and we will absorb it in definition of the angular component of the gauge field. This is the only spatial component of the gauge field that remains when the theory is dimensionally reduced to a~ring.

Since Chern-Simons theory gives a description for a two-dimensional anyon field, during the past decade there was interest to dimensionally reduce it to consistently describe one dimensional anyons. Indeed, the first dimensional reduction of Chern–Simons theory was formulated in the Hamiltonian framework and it yielded the chiral BF Hamiltonian \cite{Rabello1995}. Note that alternative reductions of the Chern-Simons Hamiltonian have also been developed, resulting in one-dimensional models different from the chiral BF one \cite{Leinaas1992,sen1994,Vathsan1998,Larisa1999,Rougerie2023,Rougerie2024,Rougerie2025,yang2025}. The Lagrangian formalism has also been employed to achieve the dimensional reduction, but it requires either the introduction of the kinetic energy for the $\mathcal{B}$ field by hand \cite{Aglietti1996,Jackiw1997}, or to couple Chern–Simons theory to a background gauge field \cite{Rojas2023,Rojas2024} in order to yield the chiral BF model. 

Here, we use the same Lagrangian formalism and introduce a boundary term to the Chern-Simons action to restore gauge invariance, which is broken by a finite spatial domain. This boundary term contains an additional degree of freedom, the $\mathcal{B}$ field, and turns out to be the BF action itself \cite{Dunne1999}. We show that the system's boundary conditions allow to match $\mathcal{B}$ with the field in the chiral boson action, with such an action resulting from the Chern-Simons one evaluated for pure gauge solutions \cite{Wen1992}. In this way, we obtain the chiral BF theory in vacuum, which can be then coupled to a one dimensional matter field. One can also couple the two-dimensional matter from the start and execute the same procedure, but an additional condition is required to obtain the matter-coupled chiral BF theory. Here, we provide the key steps of the reduction. We refer the interested reader to the Appendix A for further details.

To start our procedure, we first assign the spatial domain to be $\mathcal{D}$ and consider the (2+1)D Chern-Simons Lagrangian
\begin{align}
    \mathcal{L}_{\mathrm{CS}}&=\frac{1}{2\kappa}\varepsilon^{\mu\nu\rho}A_\mu\partial_\nu A_\rho ,\label{eq:CS_firstorder}
\end{align}
where $\varepsilon^{\mu\nu\rho}$ is the Levi-Civita tensor for  $\mu=t,r,\phi$, $A_\mu=(A^0,A_i)$, $i=r,\phi$ the covariant gauge field, and the coupling constant $\kappa$ is known as the Chern-Simons level. While this constant can have any value in classical field theory, in the quantum theory it can assume only rational or integer values, depending on matter being coupled to the field or not \cite{Iengo1991}. Here, we consider the dimensional reduction of the Chern-Simons theory to chiral BF at the classical level, so $\kappa$ is real number. In Section \ref{encoding}, however, we will see how $\kappa$ gets restricted by requesting periodicity of the phase accumulated by the wavefunction of the chiral BF theory under a closed loop.
Lagrangian \eqref{eq:CS_firstorder} is invariant only under restricted gauge transformations \cite{Wen1992}, i.e., field redefinitions $A_\mu\to A_\mu+\partial_\mu f$, where $f$ is a regular function which is zero on the boundary $f(r_0)=0$. Otherwise, the transformation of $A_\mu$ yields $\mathcal{L}_{\mathrm{CS}}\to\mathcal{L}_{\mathrm{CS}}+ D_\kappa[A_\mu,f]$, i.e., it adds a total derivative $D_\kappa$, which cannot be discarded because the boundary is not at spatial infinity. To restore gauge invariance \cite{Dunne1999}, we introduce the extension $\mathcal{L}_{\mathrm{CS + BF}}[\mathcal{B},A_\mu]=\mathcal{L}_{\mathrm{CS}}[A_\mu]-D_\alpha[A_\mu,\mathcal{B}]$,
with $\mathcal{B}$ an additional degree of freedom, and assign the transformation $A_\mu\to A_\mu+\partial_\mu f$, $\mathcal{B}\to\mathcal{B}+\alpha f/\kappa$. Then, the full action reads
\begin{equation}
S_{\mathrm{CS+BF}}=\int\mathrm{d}t\left(\int_{\mathcal{D}}\mathrm{d}\mathbf{x}\ \mathcal{L}_{CS}+\frac{1}{2\alpha}\oint_{\partial\mathcal{D}}\mathcal{B}\varepsilon^{\mu\nu}F_{\mu\nu}\right) \label{eq:S_CSBF},
\end{equation} 
where $\oint_{\partial\mathcal{D}}=r_0\int\mathrm{d}\varphi$, $\mu,\ \nu=t,\phi$, thus $F_{\mu\nu}=E_\phi$. The total Lagrangian is schematically depicted in \fref{Fig_1} b). Notice that the piece we add is the BF action itself living on $\partial\mathcal{D}$. Such a term is a total derivative for the bulk, therefore it does not alter the equations of motion (e.o.m.) for $A_\mu$ in $\mathcal{D}$, but it modifies their boundary conditions. 

To obtain the chiral BF theory on the ring, we need to integrate Chern-Simons theory in the disk along the solutions of the flux attachment condition \eqref{eq:flux_attachment_vacuum}, in order to obtain an exact boundary term which we can express as a function of $\mathcal{B}$. We consider a pure gauge configuration in the temporal gauge \cite{Wen1992}
\begin{align}
    A^0=0, \quad A_i=\partial_i\beta+\delta_{i,\phi}\frac{\omega}{r}.\label{eq:CS_puregauge_solutions}
\end{align}
Here, $\beta$ is a regular function and the second term accounts for the topological mode. We substitute the solution \eqref{eq:CS_puregauge_solutions} in the Lagrangian $\mathcal{L}_{\mathrm{CS}}$, and recast the action \eqref{eq:S_CSBF} evaluated for this solution through Stokes' theorem
\begin{equation}
\mathcal{S}_{\mathrm{CS+BF}}\overset{\mathrm{sol}}{=}\frac{1}{2}\int\mathrm{d}t\oint_{\partial\mathcal{D}}\frac{1}{\kappa}\partial_t\beta\frac{\partial_\phi}{r}\beta+\frac{1}{\alpha}\mathcal{B}\varepsilon^{\mu\nu}F_{\mu\nu},\label{eq:CS_post_int}
\end{equation}
where all fields are implicitly evaluated at $r_0$. The first term in \eqref{eq:CS_post_int} is the chiral boson action for the function $\beta$, and it is represented in \fref{Fig_1} b). Then, such a term can be recast with
the boundary condition  
\begin{equation}
\kappa\mathcal{B}(t,\phi;r_0)=\alpha\beta(t,\phi;r_0)\label{eq:boundary_condition},
\end{equation}
thus recovering the (1+1)D chiral BF theory \cite{Aglietti1996,Jackiw1997,griguolo1998,Craig2022,Rojas2023,Rojas2024} on~a ring
\begin{equation}
    \mathcal{S}_{\mathrm{CS+BF}}\overset{\mathrm{sol}}{=}\frac{1}{2}\int\mathrm{d}t\oint_{\partial\mathcal{D}}\frac{\kappa}{\alpha^2}\partial_t\mathcal{B}\frac{\partial_\phi}{r}\mathcal{B}+\frac{1}{\alpha}\mathcal{B}\varepsilon^{\mu\nu}F_{\mu\nu}\label{eq:cBF_action},
\end{equation}
after we rename the couplings as $\kappa\to\lambda$ and $\alpha\to\kappa$.

This theory can be then minimally coupled to one-dimensional matter. Alternatively, we can couple the extended Chern-Simons theory to matter in a disk, i.e., one can start from the extended Jackiw-Pi action \cite{JackiwPi}
\begin{align}    &S_{\mathrm{JP}}=S_{\mathrm{CS+BF}}+S_{\mathrm{M}}\nonumber\\&S_{\mathrm{M}}=\int\mathrm{d}t\int_{\mathcal{D}}\mathrm{d}\mathbf{x}\Psi^*(i\partial_t-A^0)\Psi+\frac{1}{2m}\Psi^*(\nabla-i\mathbf{A})^2\Psi,
\end{align}
and execute the same procedure, i.e., integrating the bulk along the equations of motion. We arrive to a chiral BF theory coupled to one-dimensional matter on the ring, together with two additional terms. The first is non-zero in the bulk and can be set to zero on the ring by requesting that the total particle number on the ring is conserved. The second term is the covariant, radial kinetic energy on the ring. It can be set to zero by requiring zero radial covariant momentum on the ring. These two requirements translate into boundary conditions for the bulk equations of motion, that are $\partial_r J_r = 0$ and $(\partial_r+iA_r)\Psi=0$, respectively, with $J_r$ the radial covariant current. Then, the integration of the Jackiw-Pi action along the e.o.m., supported by these boundary conditions, and the renaming of the couplings yields the chiral BF action
\begin{align}
    \mathcal{S}_{\mathrm{cBF}}&=\int\mathrm{d}t\oint_{\partial\mathcal{D}}   \frac{\lambda}{2\kappa^2}\partial_t\mathcal{B}\frac{\partial_\phi}{r}\mathcal{B}+\frac{\mathcal{B}}{2\kappa}\varepsilon^{\mu\nu}F_{\mu\nu}+\nonumber\\&+\!\Psi^*(i\partial_t-A^0)\Psi+\frac{1}{2mr_0^2}\Psi^*(\partial_\phi+ir_0A_\phi)^2\Psi.\label{eq:cBF_matter_action}
\end{align}

To sum up, this Section shows that the chiral BF action on a ring emerges from the gauge-invariant extension of Chern-Simons' Lagrangian on a disk, when the latter is integrated along the bulk solution and boundary conditions are prescribed.

\section{Encoded Hamiltonian derivation}\label{encoding}
In this Section, we will express the chiral BF Lagrangian coupled to a matter field in its \textit{encoded} form. With the term \textit{encoded}, we refer to a Hamiltonian formulation where the gauge degrees of freedom are expressed in terms of matter through the local conservation law. 
For instance, the gauge field in the Schwinger model \cite{Martinez2016,Muschik2017} (1D lattice model of electromagnetism) contains no dynamical degrees of freedom, thus it can be fully eliminated and gives rise to Coulomb interactions \cite{Hamer1997}. 
In the case of the chiral BF theory, the dynamical gauge degrees of freedom are absent from the Hamiltonian because the action \eqref{eq:cBF_action} is topological.
When coupled to matter, instead of Coulomb interactions, this gauge field provides statistical interactions, as shown \fref{Fig_1} b).
On a line, such interactions have the form of a coupling between density and current density, thus they are chiral. On a ring, the corresponding Hamiltonian features the same chiral interactions but in the shape of a coupling between angular current and a density-dependent magnetic flux, which are represented in \fref{Fig_1} b). Such an encoded Hamiltonian will be the target of the proposed quantum simulation with Bose-Einstein condensates.

To get the encoded Hamiltonian, we will use a procedure established by Faddeev and Jackiw \cite{FaddeevJackiw1988} to quantize constrained theories without adopting a choice of gauge. By using field redefinitions (canonical transformations) at the level of the Lagrangian and the local conservation law (Gauss' law), one eliminates the unphysical degrees of freedom. The resulting Lagrangian for the physical degrees of freedom only can be Legendre transformed into a Hamiltonian, the \textit{encoded} Hamiltonian, and the theory canonically quantized. This procedure is known to be equivalent to other quantization methods, like Dirac quantization, or to covariant quantization schemes when available \cite{Jackiw1993}. In the latter situation, eliminating the unphysical degrees of freedom from the Hamiltonian leads to the encoded Hamiltonian both in classical and quantum theories. From a quantum simulator's perspective, the use of encoded Hamiltonian is convenient because it minimizes the  degrees of freedom needed and can simplify the interactions \cite{Muschik2017,Zohar2020,Haase2021,Fontana2025}. The Fadeev-Jackiw procedure has already been used for obtaining the chiral BF Hamiltonian on a line \cite{Craig2022}. Here, the difference is that there is a topological component of the gauge potential which remains independent of matter, therefore remaining in the encoded Hamiltonian formulation. 

We start the quantization procedure from \eqref{eq:cBF_matter_action} and consider the presence of interactions $V(n)$ which depend polynomially on the local density $n = \vert \Psi \vert ^2$. From here on, time derivatives are denoted with a dot $\partial_t\Psi\equiv\dot{\Psi}$. Then, since there are no boundaries along the time and polar coordinates, we discard total derivatives with respect to these variables and write the Lagrangian in the first order formalism
\begin{align}
    \mathcal{L}_{\mathrm{cBF}}=&-A^0\!\left(n-\frac{\partial_\phi}{\kappa r_0}\mathcal{B}\right)+ \frac{\mathcal{B}\dot{A}_\phi}{\kappa}+\frac{\lambda}{2\kappa^2r_0}\dot{\mathcal{B}}\partial_\phi\mathcal{B}+\nonumber\\&+A_\phi J_\phi +i\Psi^*\dot{\Psi}+\frac{1}{2mr_0^2}\Psi^*\partial^2_\phi\Psi -V(n).\label{eq:cBF_firsorder}
\end{align}
This expression shows that the conjugate momenta $\Pi_{A_\phi}=\mathcal{B}/\kappa$, $\Pi_{\mathcal{B}}=\lambda\partial_\phi\mathcal{B}/2\kappa^2r_0 $ are not functions of the fields' time derivatives, therefore the latter cannot be expressed in terms of the momenta and the Legendre transform cannot be used. Further, in this formalism, $A^0$ acts as a Lagrange multiplier, because no space-time derivatives of this field appear. Therefore, $A^0$ can be treated as a constant multiplying the local conservation law of the theory. Independently of discarded total derivatives, the Euler-Lagrange e.o.m. of the system read
\begin{align}
    &i\dot{\Psi}+\frac{1}{2mr_0^2}(\partial_\phi+ir_0A_\phi)^2\Psi-\frac{\delta V(n)}{\delta n}\Psi=0,\\& \partial_\phi\mathcal{B} = r_0\kappa n,\label{eq:ChiralBF_flux_attachment}\\ &\dot{\mathcal{B}}=\kappa J_\phi,\label{eq:ChiralBF_eom2}\\ &r_0 \kappa E_\phi-\lambda\partial_\phi \dot{\mathcal{B}}=0.\label{eq:ChiralBF_eom3}
\end{align}
Here, $E_\phi=\dot{A}_\phi-\partial_\phi A^0$ and $J_\phi$ is the angular covariant current. The first is a Schr\"odinger equation for the $\Psi$ field under the presence of a gauge potential $A_\phi$,  the other three are differential equations for the gauge field components. 
To obtain a Hamiltonian for the independent field $\Psi$, we start the encoding with a $U(1)$ field redefinition
\begin{align}
    \Psi(t,\phi)&\to\exp\left\{i\left(-\int_{\phi_0}^\phi{\rm d}\varphi\  r_0 (A_\phi-A_\phi^T)(t,\varphi)+\right.\right.\cr
    &\left.\left.\int_{t_0}^t{\rm d}\tau\ A^0(\tau,\phi_0)-\frac{\lambda}{2\kappa}\mathcal{B}(t,\phi_0)\right)\right\}\Psi(t,\phi)\label{eq:local_gauge_1},
\end{align}
which brings the problem in a frame where the potential $\mathcal{E}=\int_{\phi_0}^\phi\mathrm{d}\varphi\ E_\phi(t,\varphi)$ is the Lagrange multiplier
\begin{align}
    \mathcal{L}_{\mathrm{cBF}}=&\mathcal{E}\!\left(n-\frac{\partial_\phi}{\kappa r_0}\mathcal{B}\right)+\frac{\lambda}{2\kappa^2r_0}\dot{\mathcal{B}}\partial_\phi\mathcal{B}-\frac{\lambda}{2\kappa}\dot{\mathcal{B}}(t,\phi_0)n+\cr&+i\Psi^*\dot{\Psi}+\frac{1}{2mr_0^2}\Psi^*(\partial_\phi+i r_0A_\phi^T)^2\Psi -V(n)\label{eq:cBF_frame1},
\end{align}
while the local conservation law \eqref{eq:ChiralBF_flux_attachment} stays the same. 
The transformation \eqref{eq:local_gauge_1} has the effect of decoupling the Lagrangian in a dynamical part, given by the chiral boson action and Schr\"odinger Lagrangian, and the conservation law of the theory, which is multiplied by $\mathcal{E}$. Also, we cannot absorb the solenoid solution $A_\phi^T$ \eqref{eq:TAP} in the field redefinition without executing a \textit{large} gauge transformation which changes the winding sector of the theory.
Next, we notice that the local conservation law \eqref{eq:ChiralBF_flux_attachment} implies
\begin{equation}
    \mathcal{B}(t,\phi)=\mathcal{B}(t,\phi_0)+\kappa \int_{\phi_0}^{\phi}r_0\mathrm{d}\varphi\ n(t,\varphi)\label{eq:CBF_fluxattachment_solution}.
\end{equation}
By substituting Eqs. \eqref{eq:TAP} and \eqref{eq:CBF_fluxattachment_solution} in Eq. \eqref{eq:cBF_frame1}, we find
\begin{align}
    \mathcal{L}_{\mathrm{cBF}}=&i\Psi^*\left(\partial_t-i\frac{\lambda}{2}\int_{\phi_0}^\phi r_0\mathrm{d}\varphi\ \dot{n}(t,\varphi)\right)\Psi+ \cr & \ \ +\frac{1}{2mr_0^2}\Psi^*(\partial_\phi+i\omega)^2\Psi-V(n).
\end{align}
A final Jordan-Wigner transformation
\begin{equation}\label{eq:JordanWigner_toH}
    \Psi(t,\phi)\to\exp{\big(i\frac{\lambda r_0}{2} \int_{\phi_0}^{\phi}\mathrm{d}\varphi\ n(t,\varphi)\big)}\Psi(t,\phi),
\end{equation}
brings the Lagrangian in canonical form
\begin{align}
&\mathcal{L}_{\mathrm{cBF}}=i\Psi^*\dot{\Psi}-\mathcal{H}^{\mathrm{enc}}_{\mathrm{cBF}},\label{eq:cBF_L_canonical}\\ & \mathcal{H}^{\mathrm{enc}}_{\mathrm{cBF}}=-\frac{1}{2m}\Psi^*\left(\frac{\partial_\phi+i\omega}{r_0}+i\frac{\lambda}{2}n\right)^2\Psi+V(n)\label{eq:cBF_Hamiltonian},
\end{align}
where the Hamiltonian can be identified.

Notice that the gauge transformations \eqref{eq:local_gauge_1} and \eqref{eq:JordanWigner_toH} change the boundary conditions of $\Psi$ throughout the encoding. That is, if the matter field in \eqref{eq:cBF_firsorder} obeys periodic boundary conditions, then the one in \eqref{eq:cBF_Hamiltonian} generally obeys twisted boundary conditions $\Psi(0)=e^{i\mathcal{A}}\Psi(2\pi)$, with 
$\mathcal{A}$ the phase accumulation. Such boundary conditions are not desirable for an experimental proposal. However, we are free to push the twisted conditions back into the original Lagrangian to absorb such a phase accumulation. To trace this back, we can write the total phase redefinition as a function of the density and initial conditions only. Indeed, the azimuthal component which appears in \eqref{eq:local_gauge_1} equals to $A_\phi-A^T_\phi=\lambda n$, which results by composing the e.o.m.\eqref{eq:ChiralBF_flux_attachment} and \eqref{eq:ChiralBF_eom3} in the temporal gauge. Likewise, $A^0$ in \eqref{eq:local_gauge_1} is zero in such a gauge choice. Therefore, the total phase that relates the matter field in \eqref{eq:cBF_firsorder} and the one in \eqref{eq:cBF_Hamiltonian} is 
\begin{equation}
    \mathcal{A}(t,\phi)=-\int_{\phi_0}^{\phi}r_0\mathrm{d}\varphi\  \frac{\lambda}{2}n(t,\varphi)-\frac{\lambda}{2\kappa}\mathcal{B}(t,\phi_0). \label{eq:Accumulated_phase}
\end{equation}
Therefore, the phase jump is $\mathcal{A}(t,2\pi)-\mathcal{A}(t,0) \equiv \mathcal{A}=-\lambda N/2$, with $N=\int_0^{2\pi} r_0\mathrm{d}\varphi \ n(t,\varphi)$. This number is generally not a multiple of $2\pi$. The accumulated phase $\mathcal{A}$ thus reflects the nontrivial gauge connection of the chiral BF theory on a ring. As a further consequence of the change of phase \eqref{eq:Accumulated_phase}, the density-dependent gauge field in \eqref{eq:cBF_Hamiltonian} is equal to $\lambda n /2$ and corresponds to half of the field in the Lagrangian \eqref{eq:cBF_firsorder}, i.e. $A_\phi-A^T_\phi=\lambda n$. As a final remark, to ensure that observables are well-defined everywhere in the quantum theory \cite{Dirac1931}, we must require that the phase $\lambda N/2 + 2\pi \omega$ accumulated by $\Psi$ around a close loop is a multiple of $2\pi$. Since the topological winding $\omega \in \mathbb{Z}$, this condition restricts the values of $\lambda$, i.e., the Chern-Simons level $\kappa$ introduced in Section \ref{ChiralBF_theory}, to obey the relation $\lambda=4\pi k /N$, with $k\in\mathbb{Z}$. Such a condition also allows the phase accumulated through the encoding procedure to be a multiple of $2\pi$. However, this restriction can be safely neglected in the limit of large particle numbers $N$, which is coherent with the numerics of Section \ref{sec:numerics}, where $\lambda$ is treated as a continuous parameter.

A key result of having the theory on a ring is the presence of a density-dependent magnetic flux, for the mere fact that the gauge potential is now azimuthal. The total flux per particle is given by
\begin{equation}
\tilde{\omega}=\int^{2\pi}_0r_0^2\mathrm{d}\varphi \Psi^*\left(\frac{\lambda}{2} n+\frac{\omega}{r_0}\right)\Psi=\frac{r_0\lambda \langle n\rangle}{2} +\omega\label{eq:magnetic_flux_variable}.
\end{equation}

The detailed form of the chiral BF encoded Hamiltonian is obtained by expanding the covariant derivative 
\begin{align}
\mathcal{H}^{\mathrm{enc}}_{\mathrm{cBF}}\!&=\!-\frac{1}{2mr_0^2}\Psi^*\partial_\phi^2\Psi\!+\!\left(\!\frac{\omega}{r_0}\!+\!\frac{\lambda n}{2}\!\right)j_\phi\nonumber\\&+\!\frac{n}{2m}\!\left(\!\frac{\omega}{r_0}\!+\!\frac{\lambda n}{2}\!\right)^{\!2}+V(n),\label{eq:cBF_Hamiltonian_exp}    
\end{align}
where $j_\phi=\mathrm{Re}\left(\Psi^*(-i\partial_\phi)\Psi\right)/mr_0$ is the angular current and $V(n)$ is an interaction term polynomial in density. The main property of \eqref{eq:cBF_Hamiltonian_exp} is the chiral current-density term proportional to $\lambda$, as all other terms coming from the covariant derivative (i.e., the two-body and three-body terms scaling as $\lambda \omega$ and $\lambda^2$ respectively) could be absorbed in $V(n)$ because of its definition, while the three-body interactions could also be discarded in the low-density regimes. At the end of next Section, we will leverage the freedom in the definition of $V(n)$ to match the chiral BF Hamiltonian \eqref{eq:cBF_Hamiltonian_exp} with an experimentally feasible Bose gas Hamiltonian.
  
Hamiltonian \eqref{eq:cBF_Hamiltonian_exp} can be quantized canonically by promoting $\Psi$ to a field operator with bosonic commutation relations, so that the computation of the quantum Heisenberg equation of motion is equivalent to the classical Euler-Lagrange e.o.m. of \eqref{eq:cBF_L_canonical}. Notably, the quantized version of the kinetic energy in the Hamiltonian \eqref{eq:cBF_Hamiltonian} is connected to a free Hamiltonian system of anyonic field operators:
\begin{align}
    &\hat{\mathcal{H}} = -\frac{1}{2mr_0^2}\hat{\eta}^\dagger\partial_\phi^2\hat{\eta},\\ &\hat{\eta}(\phi_1)\hat{\eta}^\dagger(\phi_2)-e^{i\frac{\lambda r_0}{2}\mathrm{sign}(\phi_2-\phi_1)}\hat{\eta}^\dagger(\phi_2)\hat{\eta}(\phi_1)=\delta(\phi_2-\phi_1),\label{eq:Anyon_H}
\end{align}
via the Jordan-Wigner transformation
\begin{equation}
    \hat{\Psi}(t,\phi)=e^{-i\int^{\phi}_{\phi_0}r_0\mathrm{d}\varphi \hat{A}_\omega}\hat{\eta}(t,\phi),\label{eq:JordanWigner_toA}
\end{equation}
where $\hat{A}_\omega=\omega/r_0+\lambda\hat{n}/2$. From this last expression, we notice that $\lambda \hat{n}/2$ relates both the boundary conditions of $\hat{\Psi}$ and $\hat{\eta}$ as well as their total angular momenta. Therefore, if the Bose field is periodic and with integer-valued angular momenta, the corresponding anyonic field can obey twisted boundary conditions and have a fractionalized angular momenta depending on the particle number. Importantly, the strength $\lambda$ of the chiral interaction for the Bose field controls the statistical angle of one-dimensional anyons in the continuum, and coincides with twice the sine of the lattice anyon statistical angle \cite{Bonkhoff2020}. 

To summarize, this Section shows that the chiral BF Lagrangian on a ring can be encoded into a Hamiltonian for a matter field coupled to a density-dependent and a topological gauge field. 

\section{Chiral BF on ring-shaped BECs}\label{experiment}
The second-quantized form of the chiral BF Hamiltonian \eqref{eq:cBF_Hamiltonian} describes a bosonic field coupled to a density-dependent gauge field with a winding number. Since the theory involves only local interactions, ultracold atoms can provide a suitable platform for its quantum simulation. 
Indeed, the first mean-field proposal was given by Edmonds \textit{et al.} \cite{Edmonds2013} for both linear and ring geometries. Their approach is based on a two-component Raman-coupled Bose-Einstein condensate with synthetic spin-orbit coupling, where the full system is described by the lowest eigenstate of the atom-photon system as a position-dependent dressed state. In that framework, a density-dependent gauge field emerges from contact interactions, which act as a many-body detuning of this eigenstate.

In this Section, we leverage the same experimental platform but follow a more powerful approach based on the microscopic momentum picture initially developed in \cite{Kurn2004,Spielman2009} and further refined in \cite{Craig2022}. This method allows us to derive the mapping between the chiral BF Hamiltonian and the Bose gas Hamiltonian directly at the quantum level, before any mean-field approximation. In this momentum-space picture, the coupling to the density-dependent gauge field is more naturally understood in terms of chiral interactions. At the same time, the mapping holds for lower values of the Raman coupling than in \cite{Edmonds2013}, which is crucial for reducing atom loss due to photon scattering and, thus, enable experimental realizations.

Specifically, we consider two electronic states within the lowest lying hyperfine manifold of a Bose gas that we label $\sigma=\uparrow,\downarrow$. The system is described by the bosonic field operator $\hat{\Psi}_\sigma=[\hat{\Psi}_{\uparrow}\ \hat{\Psi}_{\downarrow}]^{\mathrm{T}}$. The two internal states are Zeeman-splitted by a bias magnetic field $\mathbf{B} = B_0\hat{x}$, and further coupled by a pair of Laguerre-Gauss and Gaussian laser beams copropagating along the $\hat{z}$ axis, as shown in \fref{Fig_1} d). The Laguerre-Gauss beam carries an  orbital angular momentum $2\hbar\ell$, where $\ell$ is the quantum of  angular momentum which is transferred to the gas. A sketch of the two-level system is depicted in \fref{Fig_1}  c). Differently from the experimental realization on the line \cite{Frolian2022}, we conveniently choose the Raman transition to be red-detuned to both $D1$ and $D2$ lines, so that the same pair of beams can be employed to generate a scalar ring potential of radius $r_0$. We also add a light-sheet potential for box-like confinement along the vertical $z$ direction, with walls at $\pm z_T$. We name the total potential $V_{\mathrm{T}}$. For $-z_T<z<z_T$, it takes the form $V_{\mathrm{T}}=M\omega_r^2 (r-r_0)^2/2$, where $M$ is the atomic mass and $\omega_r$ the radial trapping frequency. The gas takes the shape of a hollow cylinder, as depicted in \fref{Fig_1}  d). Its non-interacting properties are described in the laboratory frame by the one-body Hamiltonian \cite{Chen20181}
\begin{align}
&\hat{H}_{0}=\int\rm{d}\mathbf{x}\sum_{\sigma_1,\sigma_2}\hat{\Psi}^\dagger_{\sigma_1}(\mathbf{x})\mathcal{H}_{0,\sigma_1,\sigma_2}\hat{\Psi}_{\sigma_2}(\mathbf{x}),\label{eq:H0_allspace}\\
&\mathcal{H}_{0}=\left(-\frac{\hbar^2}{2M}\nabla^2+V_{\mathrm{T}}\right)\otimes\mathbb{I}+\cr&\quad -\frac{\hbar\delta}{2}\sigma_z+\frac{\hbar\Omega}{2}\left(e^{+i 2\ell \phi}\sigma_++e^{- i2\ell \phi}\sigma_-\right)
\label{eq:H0}.
\end{align}
Here, $\mathbf{x}$ labels the position in the three-dimensional space, $\phi$ is the polar coordinate, $\Omega$ and $\delta$ indicate the two-photon Rabi frequency and single-particle detuning, respectively, and $\mathbb{I}$, $\sigma_{z}$ and $\sigma_\pm=(\sigma_x\pm i \sigma_y)/2$ are the identity and Pauli matrices in spin space. Analogously to the linear case \cite{Kurn2002,Kurn2004,Spielman2011}, this Hamiltonian can be made manifestly rotationally invariant by a local spin~rotation
\begin{equation}
\hat{\Psi}\to\mathrm{diag}\left[e^{+i\ell\phi},\ e^{-i\ell \phi}\right]\hat{\Psi}\label{eq:local_spin_rotation},
\end{equation}
which leads to
\begin{align}
&\mathcal{H}_{0}=\left(-\frac{\hbar^2}{2M}\nabla^2+V_{\mathrm{T}}+\frac{\hbar^2\ell^2}{2Mr^2}\right)\otimes\mathbb{I}+\cr&\quad+\left(2\hbar\ell\frac{-i\hbar\partial_\phi}{Mr^2}-\frac{\hbar\delta}{2}\right)\sigma_z+\frac{\hbar\Omega}{2}\sigma_x\label{eq:H0p}.
\end{align}
In this rotating frame, the Hamiltonian shows the spin-orbital angular momentum coupling $L_z\sigma_z$, for $L_z=-i\hbar\partial_\phi$, plus a scalar potential $\propto r^{-2}$. To diagonalize the Hamiltonian \cite{Chen20181}, we can disregard the transverse directions, because the experimental setup involves tight confinement around $r_0$ in the radial direction and homogeneity $\hat{z}$ direction.
Since Hamiltonian \eqref{eq:H0p} commutes with $L_z$, we look for eigenstates of the type $\hat{\Psi}_\sigma(m)=e^{im\phi}\hat{\Psi}_\sigma$. Here, $m$ is the quasi-angular momentum, i.e., the angular momentum number in the rotating frame, and this form allows the substitution $-i\partial_\phi\to m$ in \eqref{eq:H0p}.

Until the end of this Section, we work in units of the recoil energy $E_\mathrm{R}=\hbar^2\ell^2/2Mr_0^2$ and length $r_0/\ell$ to simplify notation. The momentum-dependent rotation 
\begin{align}
    &\begin{bmatrix}\hat{\Psi}_\uparrow\\ \hat{\Psi}_\downarrow\end{bmatrix}(m)=\hat{R}(m)\begin{bmatrix}\hat{\Psi}_+\\ \hat{\Psi}_-\end{bmatrix}(m)\label{eq:Rdress},\\
    &\hat{R}(m)=\exp{i\frac{\theta(m)}{2}\sigma_y}=\begin{bmatrix}\cos{\frac{\theta(m)}{2}}& -\sin{\frac{\theta(m)}{2}}\\ \sin{\frac{\theta(m)}{2}}& \cos{\frac{\theta(m)}{2}}\end{bmatrix}\label{eq:Rmat},    
\end{align}
brings the Hamiltonian \eqref{eq:H0p} in diagonal form
\begin{equation}
   \mathcal{H}_0 =\sum_m \hat{R}(m) \mathcal{H}_m \hat{R}^\dagger(m),\quad\quad \mathcal{H}_m \hat{\Psi}_\pm = E_\pm \hat{\Psi}_\pm.
\end{equation}
Here, we introduced the rotation angle $\theta(m)=\tan^{-1}{\Omega/\tilde{\delta}}$, with $\tilde{\delta}=\delta-4m\ell/r^2$, as well as the dressed basis $\hat{\Psi}_{\pm}$ and dressed bands
\begin{equation}
    E_\pm\left(\frac{m}{r}\right)=\frac{m^2+\ell^2}{r^2}\pm\frac{1}{2}\sqrt{\Omega^2+\left(\delta-\frac{4m\ell}{r^2}\right)^2}\label{eq:SPLB},
\end{equation}
from which we identify the band gap $\tilde{\Omega}=\sqrt{\Omega^2+\tilde{\delta}^2}$. Notice that for every (quasi) angular momentum number $m$, we introduce a family of momenta $m/r$. Similarly, there is a family of Raman momenta $\ell/r$ with value of unity when $r=r_0$. These bands have a spin composition that is momentum dependent through a spin polarization parameter $P(m)=\tilde{\delta}/\tilde{\Omega}$, a property that is depicted with the bands spin texture in \fref{Fig_1} e). 
When the interaction energy is small with respect to the band gap, we can truncate the Hamiltonian to its lower band. We can also Taylor expand this lower band with respect to the quasi-angular momentum by writing $m/r=(m_0+q)/r$, with $m_0$ the center of the expansion. The expansion stays accurate at low order as long as $m_0$ is close to the center of mass momentum and if the momentum spread of states we consider is not too large. Plus, we only need one center of expansion if the band has only one minimum, which is true when $\Omega>4$. Up to third order in $q\ell/r^2\tilde{\Omega}$, we find
\begin{equation}
    E_-\left(\frac{q}{r}\right)\!=\!E_-\left(\frac{m_0}{r}\right)+\!\frac{\left(q/r-A_S\right)^2}{M^*}\!+W_0+O\left(\frac{\ell q}{r^2\tilde{\Omega}}\right)^4.
\end{equation}
We then interpret the dressed state as a particle with a larger effective mass $M^*$ under a synthetic vector potential $A_S$, of the form
\begin{align}
    &M^*=\left(1-4\left(\frac{\ell}{r}\right)^2\frac{\Omega^2}{\tilde{\Omega}^3}\right)^{-1},\\
    &A_S=-M^*\left(\frac{m_0}{r}+\frac{\ell}{r}\frac{\tilde{\delta}}{\tilde{\Omega}}-8\left(\frac{\ell}{r}\right)^3\frac{\Omega^2\tilde{\delta}}{\tilde{\Omega}^5}\frac{q^2}{r^2}\right)\label{eq:AS},
\end{align}
as depicted in \fref{Fig_1}  e). The particle also feels and additional scalar potential $W_0=-A_S^2/M^*$. These results coincide with the ones for linear spin-orbit coupling \cite{Craig2022} for $m_0/r\to k_0$, $\ell/r\to k_R$, $q/r\to q$. The static gauge potential cannot be gauged away since it is a space-dependent function and its reabsorption can lead to a multivalued field  \cite{Kleinert2008}. Instead, we keep it, as it maps into the solenoid solution \eqref{eq:TAP} as we shall see later. We can write the Taylor-expanded lower-band Hamiltonian back in position space by applying the inverse Fourier transform $\mathrm{FT}^{-1}$ as $\hat{H}_0=\int\mathrm{d}\mathbf{x}\ \mathrm{FT}^{-1}(\hat{\Psi}_-^\dagger E_-\hat{\Psi}_-)(\mathbf{x})$. This corresponds to taking the dressed field back to position space and replacing $q\to-i\partial_\phi$, so that we obtain
\begin{align}
    &\hat{H}_0=\int\mathrm{d}\mathbf{x} \Bigg[ \hat{\Psi}_-^\dagger\bigg(E_{-}\left(\frac{m_0}{r}\right)\!-\nabla^2_\perp-\frac{\partial_\phi^2}{r^2M^*}+\cr&\quad+O\left(\frac{-i\ell \partial_\phi}{r^2\tilde{\Omega}}\right)^4\bigg)\hat{\Psi}_-+A_S(m_0)\hat{j}_\phi\Bigg]\label{eq:sp_Hlab},
\end{align}
and we recognize the definition of the normal ordered current operator $\hat{j}_\phi=(\hat{\Psi}_-^\dagger\partial_\phi\hat{\Psi}_--(\partial_\phi\hat{\Psi}_-)^\dagger\hat{\Psi}_-)/(i M^*r)$. This concludes the treatment of the non-interacting term.

\begin{figure*}[t!]
\centering
\includegraphics[width=1.0\linewidth]{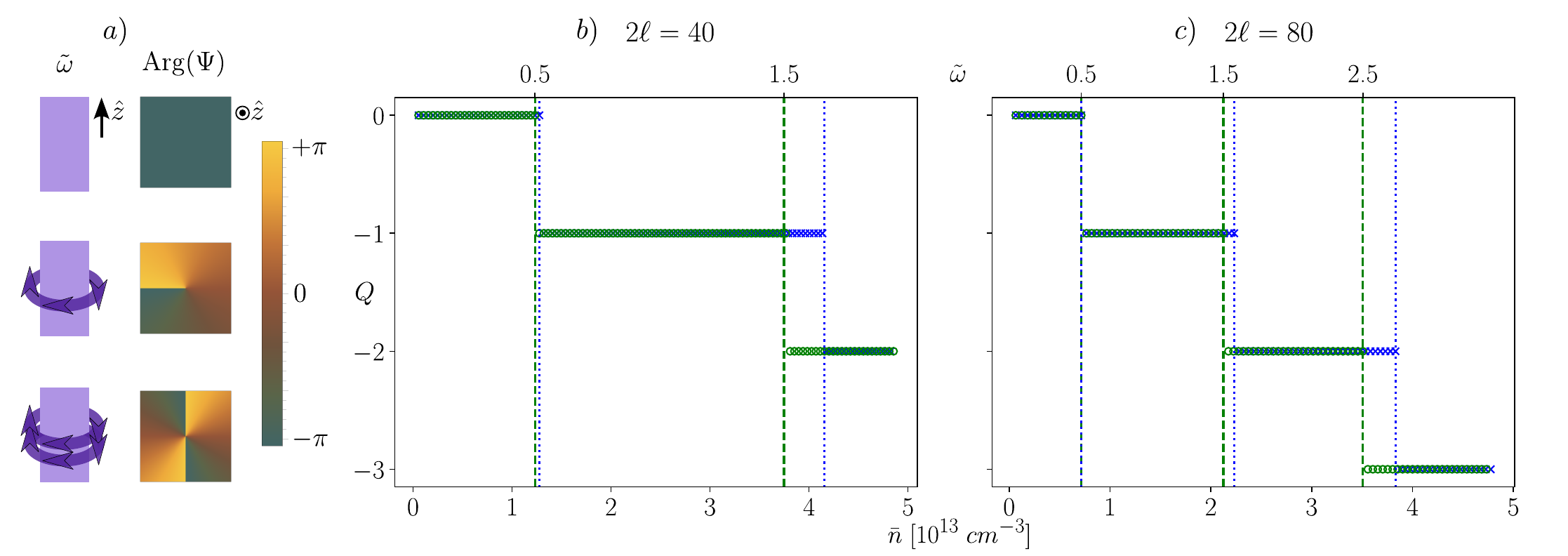}
\caption{Density-dependent angular momentum. a) The density-dependent magnetic flux  $\tilde{\omega}$ is depicted as a solenoid extended in the $\hat{z}$ direction. When its value increases by a unit, which is depicted as an anti-clockwise current in the $xy$-plane, the ground state picks up a phase winding in the same plane, so that its circulation follows the opposite direction of the current. b)- c)  The  angular momentum $Q$ as a function of the mean density $\bar n$ in the ground state for both the chiral BF \eqref{eq:H_lab_effective} (green circles) and the two-component Raman-coupled \eqref{eq:Hlab} (blue crosses) models. We scan it by varying the particle number for  fixed trapping configurations with two different Laguerre-Gauss modes $2\ell=40$(b), $80$(c). The points where the momentum changes are marked with vertical dotted blue (dashed green) lines for the Raman-coupled two-component model (chiral BF model), and they are measured in terms of $\tilde{\omega}$ \eqref{eq:magnetic_flux_variable}.}
\label{Fig_2}
\end{figure*}
To transform the single particle Hamiltonian \eqref{eq:sp_Hlab} into the chiral BF Hamiltonian 
\eqref{eq:cBF_Hamiltonian_exp} we have to couple the bosons to the density-dependent field  $\hat A_\omega$. In fact, for small $\lambda$, this coupling reduces to current-density and density-density interactions. While the latter are naturally present in a condensate, the former arise in a Raman-coupled Bose gas with imbalanced interactions \cite{Craig2022}, as we show below. We first regard the single particle Hamiltonian as the dominant term and density-density interactions as perturbation
\begin{equation}
   \hat{H}_{\mathrm{lab}}=\hat{H}_{0}+\hat{H}_{\mathrm{int}}.\label{eq:Hlab}
\end{equation}
In this way, we restrict to the lowest energy band, and at leading order project the s-wave interactions on the corresponding basis, the dressed basis. To this end, we take the condensate $s$-wave contact interactions 
\begin{equation}
\hat{H}_{\mathrm{int}}=\frac{1}{2}\int\mathrm{d}\mathbf{x}\hat{\Psi}^\dagger(\mathbf{x})\begin{bmatrix}g_{\uparrow\uparrow}\hat{\Psi}^\dagger_{\uparrow}\hat{\Psi}_{\uparrow}&g_{\uparrow\downarrow}\hat{\Psi}^\dagger_{\uparrow}\hat{\Psi}_{\downarrow}\\ g_{\downarrow\uparrow}\hat{\Psi}^\dagger_{\downarrow}\hat{\Psi}_{\uparrow}&g_{\downarrow\downarrow}\hat{\Psi}^\dagger_{\downarrow}\hat{\Psi}_{\downarrow}\end{bmatrix}\hat{\Psi}(\mathbf{x}),\label{eq:contact_interactions_positionspace}
\end{equation}
express them in quasi-momentum space
\begin{align}
&\hat{H}_{\mathrm{int}}=\frac{1}{2}\int\prod_i\mathrm{d}\mathbf{k}_{i\perp}\sum_{\sigma_{1},\sigma_{2},\{m_i\}_1^4}g_{\sigma_1,\sigma_2}\times\cr&\delta(m_4+m_3-m_2-m_1)\hat{\Psi}_{\mathbf{k}_4\sigma_2}^\dagger\hat{\Psi}_{\mathbf{k}_3\sigma_1}^\dagger\hat{\Psi}_{\mathbf{k}_2\sigma_2}\hat{\Psi}_{\mathbf{k}_1\sigma_1},\label{eq:contact_interactions_momentumspace}
\end{align}
where $\mathbf{k}_i=m\hat{\mathbf{k}}_\phi+\mathbf{k}_\perp$,
and apply the transformation in \eqref{eq:Rdress} and \eqref{eq:Rmat}. Under the same assumptions used to manipulate the non-interacting term, we first truncate the interactions to the term describing scattering in the lower band only:
\begin{align}
    &\hat{H}_{\mathrm{int}}\approx\frac{1}{2}\int\prod_i\mathrm{d}\mathbf{k}_{i\perp}\sum_{\{m_i\}_1^4}g_{-}\times\cr&\delta(m_4+m_3-m_2-m_1)\hat{\Psi}_{\mathbf{k}_4-}^\dagger\hat{\Psi}_{\mathbf{k}_3-}^\dagger\hat{\Psi}_{\mathbf{k}_2-}\hat{\Psi}_{\mathbf{k}_1-},
\end{align}
where 
\begin{equation}
    g_{-}=g_{\uparrow\uparrow}S_{1-4}+g_{\downarrow\downarrow}C_{1-4}+g_{\uparrow\downarrow}(S_1S_3C_2C_4+C_1C_3S_2S_4)\label{eq:g_lower},
\end{equation}
with
\begin{align}
&S_i=\sin{\frac{\theta(m_i)}{2}},\quad C_i=\cos{\frac{\theta(m_i)}{2}}\label{eq:polarization_coeff},\\&S_{1-4}=\prod_{i=1}^4S_i,\quad C_{1-4}=\prod_{i=1}^4C_i.
\end{align}
Notably, the two-body coupling that describes the intra-lower band interactions depends on both the angular-momentum and the Raman parameters. This property confers a degree of tunability to the dressed state interactions, in analogy with previous works on different geometries \cite{Williams2012,Josep2021DPSS,Josep2021ESQPT,Frolian2022IC,Frolian2022,Craig2022}.
We can expand the lower band scattering strength \eqref{eq:g_lower} up to first order in powers of $q\ell/r^2\tilde{\Omega}$ around $m_0/r$ to find 
\begin{equation}
    g_{-}(\{q_i\};m_0)=g_{0}(m_0)+\frac{\sum_i\ell q_i}{r^2\tilde{\Omega}}g_{1}(m_0)+O\left(\frac{\ell q_i}{r^2\tilde{\Omega}}\right)^2,
\end{equation}
where
\begin{align}
   &g_{0}(m_0)=\frac{1}{4}\!\left(g_{\uparrow\uparrow}(1+P)^2\!+g_{\downarrow\downarrow}(1-P)^2\!+2g_{\uparrow\downarrow}(1-P^2)\right),\\&g_{1}(m_0)=\frac{1}{2}\left(\frac{4\tilde{\delta}}{\tilde{\Omega}}g_{0}+g_{\downarrow\downarrow}(1-P)^2-g_{\uparrow\uparrow}(1+P)^2\right),
\end{align}
with $P=P(m_0)$. Finally, we transform the Taylor expanded interactions back to position space, and collect them together with the non-interacting term to find the effective atomic Hamiltonian
\begin{align}
&\hat{\mathcal{H}}^{\mathrm{eff}}_{\mathrm{lab}}=\hat{\Psi}^\dagger_-\bigg(\mathcal{E}_-\left(\frac{m_0}{r}\right)-\nabla^2_\perp-\frac{\partial_\phi^2}{r_0^2M^*}\bigg)\hat{\Psi}_-+\cr&\quad+\frac{1}{2}g_{0}\hat{n}^2+:\hat{j}_\phi\left(A_S(m_0)+\frac{\lambda}{2}\hat{n}\right):\label{eq:H_lab_effective}
\end{align}  
Here, $\lambda=2M^*g_{1}/\Omega$ is the coupling between current $\hat{j}_\phi$ and density $\hat{n}=\hat{\Psi}^\dagger_-\hat{\Psi}_-$ and $:\ :$ denotes the normal ordering. In \fref{Fig_1} e), this current-density interaction is depicted at the mean-field level by a density-dependent displacement of the lower band, which arises from imbalanced interactions between bare spin states. As shown in \fref{Fig_1} f), these interactions are manifest in the density-dependent shift of the angular momentum in the ground state and in having different left and right speed of sound (see Section \ref{sec:numerics}). The Hamiltonian now displays the same density-dependent gauge potential $\hat{A}_\phi=\lambda\hat{n}/2$ and topological one $\omega/r_0=A_S(m_0)$ as defined in the chiral BF Hamiltonian \eqref{eq:cBF_Hamiltonian_exp} we obtained in the previous section, with the distinction that $r_0A_S$ is not bounded to integer values while $\omega$ is. Comparing this lower-band Hamiltonian with the chiral BF one \eqref{eq:cBF_Hamiltonian_exp}, we see that the two models are equivalent up to powers of the density squared. The effective three-body term $\propto n^3$, which comes from expanding the covariant derivative in chiral BF Hamiltonian \eqref{eq:cBF_Hamiltonian_exp} but absent in the lower-band one \eqref{eq:H_lab_effective}, can be absorbed in the definition of $V(n)$ of the former Hamiltonian, because $V(n)$ is allowed to be any polynomial in the density.  Even if this was not the case, the strength of three-body interactions is negligible in the low-density regime considered in this work, that is 
\begin{equation}
    \hat{\mathcal{H}}_{\mathrm{cBF}}=\hat{\mathcal{H}}^{\mathrm{eff}}_{\mathrm{lab}}+\mathcal{O}(\lambda^2n^3).
\end{equation} Therefore, to our level of approximation the Raman-dressed gas experiences the same density-dependent flux defined for the chiral BF theory~\eqref{eq:magnetic_flux_variable}.

\section{Properties of the Raman-dressed gas}\label{sec:numerics} 
The mapping that links the two-component Raman-coupled gas to the chiral BF model is based on two approximations. Namely, we require that the band gap is large compared to the interband scattering processes, and that the momentum spread of the wave-packet is not too large. In this Section, we check the predictions of the lower-band Hamiltonian \eqref{eq:H_lab_effective} at the level of experimental observables. We do so with numerical simulations of the mean-field equations of the dressed gas \eqref{eq:Hlab} and of the lower-band Hamiltonian \eqref{eq:H_lab_effective}. Since we are interested in the low-density regime in which tree-body losses in a Bose mixture are negligible, we study the Hamiltonians in the mean-field limit by solving Gross-Pitaevskii equations.

We perform effective two-dimensional simulations using the XMDS package \cite{Dennis2013}. To reduce the dimensionality of the problem, we remove the radial direction and rescale the two-body interaction couplings with the mean transverse density. Such a mean density is given by $(\sqrt{2\pi}r_0 a_r)^{-1}$, assuming a gaussian profile with standard deviation $a_r =\sqrt{\hbar/M\omega_r}$, with $\omega_r$ the radial trapping frequency. This assumption is well justified given the strong transverse confinement we consider. The simulations are run using an adaptive-step Runge-Kutta algorithm in a rectangular domain $[-\pi,\pi ]\times [0,z_0]$, corresponding to the $\hat{\phi}$ and $\hat{z}$ directions, the latter with boundary $z_0>z_T$. We use a $512\times 32$ grid with periodic boundary conditions along $\hat{\phi}$ and mirror symmetry along $\hat{z}$.

As parameters for the simulations, we compute the two-photon effective couplings as detailed in \cite{Wei2013} for $\mbox{}^{39}$K at a bias magnetic field $B_0 = 386 \ \mathrm{G}$, where the $s$-wave scattering lengths are $a_{\uparrow\uparrow}/a_0=-2.2$, $a_{\downarrow\downarrow}/a_0=63.1$, $a_{\uparrow\downarrow}/a_0=-10.7$. We set the wavelength of the Raman beams to $773 \ \mathrm{nm}$, and choose their frequency difference to have $\delta=0$. 
We investigate two Laguerre-Gauss modes $2\ell=40,\ 80$ and compare their results by keeping the relevant experimental parameters matched between the two situations. Specifically, we consider different laser powers for the two Laguerre-Gauss modes in order to achieve the same radial trapping frequency $\omega_r/2\pi\approx 0.5\ \mathrm{kHz}$ and the same Rabi frequency $\Omega/2\pi\approx2.9\ \mathrm{kHz}$ at the trap centers $r_0\approx w\sqrt{\ell}=35.7,\ 50.6\ \mathrm{\mu m}$ defined by the two modes, with $w=8\ \mathrm{\mu m}$ the Laguerre-Gauss laser beam waist. Furthermore, in order to achieve the same mean densities at a given particle number for both Laguerre-Gauss modes, we compensate the change of radius $r_0$ by adjusting the box confinement along the $\hat{z}$-axis to a distance of $z_T\approx 30,\ 21\ \mathrm{\mu m}$ for $2\ell=40,\ 80$, respectively. We will keep this choice of parameters throughout the remainder of the Section, and scale them in the system's natural units for the numerical simulations. As a final remark, we notice that $\lambda$ enters in the Hamiltonians \eqref{eq:cBF_Hamiltonian_exp}\eqref{eq:H_lab_effective} only through its product with the density. Since lambda can be always taken positive by properly labelling the two atomic states as spin up and down, it is physically equivalent, and more convenient experimentally, to vary the mean density rather than $\lambda$. Thus, we benchmark numerically this situation.

\subsection{Quantized currents}\label{sec_Flux}
 
The first key property of the Laguerre-Gauss dressed gas in a ring geometry we examine is that it sustains self-generated persistent currents in its ground state, i.e., no external potential is required to achieve them. In order to characterize these currents, we quantify the dependency of the ground state angular momentum on the density in terms of the magnetic flux $\tilde{\omega}$ \eqref{eq:magnetic_flux_variable} arising from the encoded Hamiltonian \eqref{eq:cBF_Hamiltonian}. We use imaginary time evolution of the Gross-Pitaevskii equation for both the chiral BF model \eqref{eq:H_lab_effective} and the Raman-dressed gas \eqref{eq:Hlab} to retrieve the ground state of the systems in both situations. From this, we compute the angular momentum $Q$, i.e., the mean-field phase winding, as well as the magnetic flux $\tilde{\omega}$. 

We illustrate their relation in \fref{Fig_2}.
In the chiral BF model \eqref{eq:cBF_Hamiltonian}, every time that $\tilde{\omega}$, which is a continuous quantity, is increased by a full unit, the ground state angular momentum, which is quantized and integer-valued, also changes by a unit. The system current circulates in a direction which is opposite to the one of the ideal external currents which would generate the magnetic flux, as depicted in \fref{Fig_2} a). In this way, the chiral gas effectively screens the magnetic flux that is piercing through the ring.
As this flux is density-dependent, we detail this relation in \fref{Fig_2} (b)-(c) for zero $A_S$ \eqref{eq:AS}, where we show $Q$ as a function of the mean density 
\begin{equation}
    \bar n= \frac{2 N}{\sqrt{2\pi}r_0a_r}\int_0^{z_{0}}\int_{-\pi}^{\pi}\mathrm{d}z\mathrm{d}\varphi\vert\Psi(\varphi,z)\vert^4,
\end{equation}
where $\Psi$ is the ground state mean-field wave function normalized to unity,
for both the chiral BF model and the Laguerre-Gauss-dressed gas. Both systems show a good qualitative agreement, with the angular momentum decreasing in a step-like pattern as the density is increased. Similar to the change of momentum discussed in \cite{Cominotti2014}, we associate the jumps in $Q$ to a variation of a unit of flux, which here is density-dependent instead of being an external variable. For the chiral BF model, this flux is given by $\tilde{\omega}$ \eqref{eq:magnetic_flux_variable}, shown in the top horizontal axis figure. In the Raman-dressed gas, this behaviour is qualitatively the same and we also observe jumps in $Q$ as a function of the density. The density values where the jump in momentum takes place are however shifted from those of the chiral BF in a density-dependent way. Such density-dependent corrections to the effective flux arise from two-body interactions coupling lower and higher dressed states, thus renormalizing its value. These corrections become stronger at higher density, as detailed in Appendix B by means of a perturbative calculation. Here, we note that by increasing the angular momentum carried by the Laguerre-Gauss beam, higher values of magnetic fluxes can be realized at lower densities, where the higher-band corrections are~negligible. This numerical analysis shows that the angular momentum is unchanged under substantial density variations, thus demonstrating the robustness associated with the non-trivial topology of the ring geometry and the topological nature of the chiral interactions.
\begin{figure}[htbp]
\includegraphics[width=0.99\linewidth]{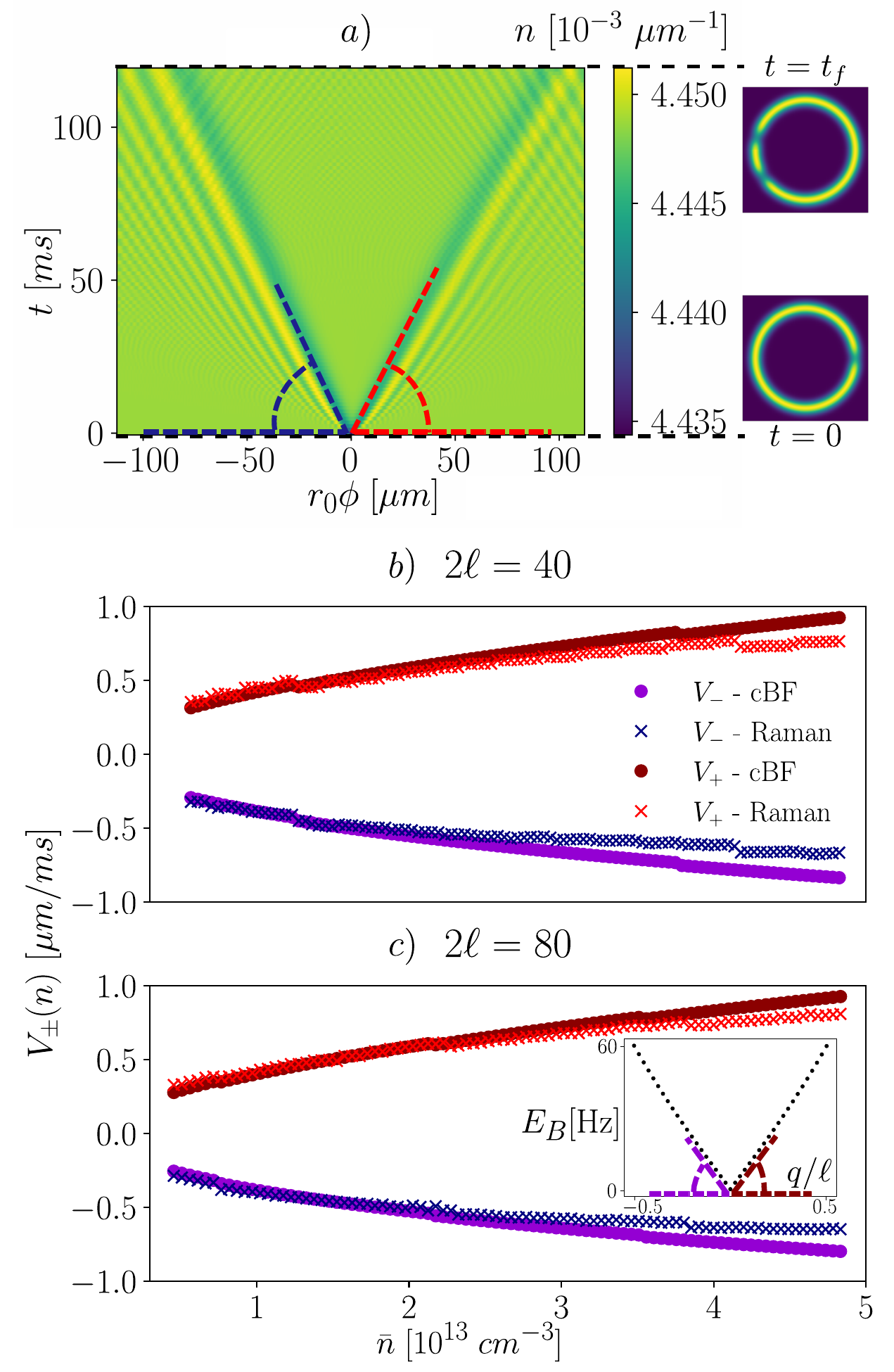}
\caption{Chiral sound velocities. a) Time evolution of the perturbed ground state spatial density profile as a function of $r_0\phi$ for $Q=-2$, and the mean density values $\bar n$ given by the colormap. The horizontal black lines link the timestamp to the ring-shaped density profile for $t=0,\ 120\ \mathrm{ms}.$ The blue and red dashed angles indicate the left and right dip velocities, respectively. b)- c) Chiral sound velocities as a function of the mean density, for two  Laguerre-Gauss modes $2\ell=40,\ 80$. The Raman-coupled two-component model \eqref{eq:Hlab} (chiral BF model \eqref{eq:H_lab_effective}) right velocity is marked with red crosses (dark red circles), whereas the left one is marked with blue crosses (violet circles). Inset: Linear part of the Bogoliubov energy spectrum of the chiral BF model \eqref{eq:H_lab_effective} as a function of the quasi-angular momentum of the perturbation. Violet (dark red) dashed angles correspond to the chiral BF left (right) sound velocities. }
\label{Fig_3}
\end{figure}
\subsection{Chiral sound velocities}\label{sec_Bogoliubov}
The second defining property of the system that we investigate is the breaking of Galilean invariance due to the chiral interactions. As a footprint of this property, chiral asymmetric expansion and chiral soliton formation were observed in the experiment on the linear geometry \cite{Frolian2022}, as expected from the predictions given by the chiral BF theory. Here, we explore the absence of Galilean invariance by perturbing the ground state of the system with an external potential that creates a density dip, and then analyzing the modes which are produced when the potential is removed. In the simulations, we model this perturbative potential with a gaussian profile having repulsive strength $V=0.01\  E_R$, focused with a waist of $1\ \mathrm{\mu m}$ at $\phi=0$, which we add to the mean-field equations of systems \eqref{eq:H_lab_effective} and \eqref{eq:Hlab}. After finding the ground state of both of the chiral BF and Raman-dressed perturbed Hamiltonians, we let the state evolve in real time in the absence of the potential.
As seen in \fref{Fig_3} a), the perturbation splits in two dips traveling in opposite directions with different velocities, i.e., they are chiral. We connect these velocities to the speed of sound obtained using Bogoliubov analysis on the low-density chiral BF Hamiltonian \eqref{eq:H_lab_effective}, thus considering the perturbation to the ground state to be composed of linear excitations. For the details of the derivations, see Appendix C. As done in \cite{Arazo2023} for zero contact interactions, we obtain the energy of the linear excitations as 
\begin{align}\label{eq:Bog_E}
    &E_B(q,Q)=\frac{q(2Q-2r_0A_S+\lambda r_0 \bar n)}{M^*r_0^2}+\cr&\pm \! \sqrt{\frac{q^2}{M^*r_0^2}\left(\frac{q^2}{M^*r_0^2} \!+\! 2\left(g_0 \bar n \!+\! \frac{2\lambda Q \bar n}{M^*r_0} \!+\! \frac{\lambda^2 \bar n^2}{2M^*}\right)\right)}. 
\end{align}
Here, $Q$ is the ground state angular momentum and $q$ is the relative momentum of the excitations. From \eqref{eq:Bog_E}, we obtain the right/left group velocities as
\begin{equation}    V_{\pm}=\pm r_0\big(E_B(\pm 1,Q)-E_B(0,Q)\big)-Qr_0^{-1}\label{eq:Bog_V}.
\end{equation}
In \fref{Fig_3} b)-c), we plot the left and right velocities as a function of the mean density. We show both the velocities as extracted by the simulations of the Laguerre-Gauss-dressed gas \eqref{eq:Hlab} and those obtained from \eqref{eq:Bog_V} for the chiral BF \eqref{eq:H_lab_effective} ground state. The small discontinuities in the sound velocities occur because they depend on the ground-state angular momentum \eqref{eq:Bog_V}, which changes in discrete integer steps as the density varies, thus yielding corresponding jumps in the velocities. We observe a good agreement between models, with the velocities and the left/right asymmetry increasing with both the density and the ring radius. As discussed in the analysis of the quantized currents, the scattering processes between the two bands slightly shift the velocities of the dressed gas with respect to the effective model in a density-dependent way. This effect becomes noticeable only for densities $\bar n > 3\times10^{13}\ \mathrm{cm}^{-3}$.

\section{CONCLUSIONS}
In this work, we have proposed a scheme to realize a topological gauge theory on a manifold of non-trivial geometry. In particular, we have shown how the chiral BF theory on a ring arises naturally from Chern-Simons theory on a disk, and that the former can be implemented with optically-dressed ultracold atoms in a ring-shaped trap. The proposal leverages the encoding of gauge degrees of freedom into a matter field, so that the theory can be implemented with a two-component Raman-coupled gas.
We have quantified the interplay between the topology of the model and the topology of the supporting space by two observables, quantized current and chiral sound velocity. Together with the gas density acting as a self-generated magnetic flux, these two observables uniquely characterize the chiral BF theory on the ring. In our zero-temperature numerical analysis, we have shown that they are robust and accessible in realistic experimental conditions. We expect these signatures to be robust also at finite temperature as long as the condensate fraction of the gas is not negligible.

Our results open interesting perspectives for exploring quantum phases with ultracold atoms induced by effective chiral interactions. In particular, the interplay between current and density could allow the investigation of chiral solitons in our system, building on recent advances in the exploration of solitons in ring geometries \cite{Rabec2025}. Our setup also provides a concrete step towards engineering Chern-Simons-like physics \cite{Rojas2020,Kamal2024}, i.e., density-dependent magnetic fields, by coupling multiple real or synthetic rings \cite{BoadaCeli2012,Celi2014, Celi_synthetic2024, Mancini-15, Stuhl-15, Livi-16, Kolkowitz-17, Bouhiron_Science_2024_4D_Hall} and exploring the transport properties of such systems \cite{Fabre_PRL_2022_Laughlin, Zhou_Science_2023_Hall}. Lattice implementations of the chiral BF theory would also allow the study of strongly interacting one-dimensional anyonic matter \cite{Yajiang2009,Keilmann2011,Greschner20152,Tang2015,Strater2016,Zhang2017,Santos2018}, with synthetic dimensions providing a resource to engineer nontrivial topologies \cite{BoadaCeli2015}. Furthermore, while we have investigated here the feasibility of our proposal for potassium atoms, experiments could also be performed with other atomic species. Cesium, which has a rich Feshbach spectrum to tune interatomic interactions and a larger fine structure splitting, would allow one to implement this scheme with reduced losses from inelastic photon scattering caused by the Raman beams. 

In addition to these experimental avenues, another interesting direction is the comparison of the different 1D reductions of two-dimensional Chern-Simons theory that, depending on the prescription, lead to either chiral BF theory  \cite{Rabello1995,Aglietti1996,Jackiw1997,griguolo1998,Rojas2023,Rojas2024}, Calogero-Sutherland \cite{Vathsan1998,Larisa1999}, Tonks-Girardeau \cite{sen1994,Rougerie2023,Rougerie2024} or cubic-quintic Schrödinger theories \cite{Rougerie2025,yang2025}. A systematic study of the dependence of reductions on different boundary conditions would be worth both at the classical and quantum level \cite{Leinaas1992}.

Finally, we emphasize that the ring geometry discussed here generally does not allow phase accumulation to be reabsorbed under boundary conditions \cite{Calabrese2009}. If the phase accumulation further depends on the internal degrees of freedom of the matter, then subjecting such a system to non-commuting \cite{Ohberg2007}, density-dependent gauge fields in this nontrivial geometry could enable the realization of non-Abelian anyonic statistics. Our work provides a framework for studying anyons in quasi one-dimensional systems, and interpolates between one- and two-dimensional exotic statistics by implementing local conservation laws and without relying on strong interactions.

\begin{acknowledgments}
We thank N. Rougerie for the exchanges on the different dimensional reductions of the Chern-Simons theory, A. Pelster for fruitful discussions on Bogoliubov modes, and C. S. Chisholm for assistance with the XMDS2 package.
\textit{Funding:} We acknowledge funding from the European Union (ERC CoG-101003295 SuperComp), the Spanish Ministry of Science, Innovation and Universities MCIU/AEI/10.13039/501100011033/FEDER, UE (projects LIGAS
PID2020-112687GB-C21, MAPS PID2023-149988NB-C22 and Severo Ochoa CEX2024-001490-S at ICFO, and projects LIGAS PID2020112687GB-C22 and MAPS PID2023-149988NB-C21 at UAB), the EU QuantERA project DYNAMITE (funded by MICN/AEI/ 10.13039/501100011033 and by the European Union NextGenerationEU/PRTR PCI2022-132919 (Grant No. 101017733)), Deutsche Forschungsgemeinschaft (Research Unit FOR2414, Project No. 277974659), Generalitat de Catalunya 
(AGAUR SGR 2021-SGR-01448 at ICFO and 2021-
SGR-00138 at UAB, CERCA program), Fundació
Cellex and Fundació Mir-Puig. C. I. acknowledges support from the Spanish Ministry of Science and Innovation MCIN/AEI/10.13039/501100011033 and ESF through the predoctoral contract PRE2021-099050. J.C. acknowledges support from the Spanish Ministry of Economic Affairs and Digital Transformation through the QUANTUM ENIA project call – Quantum Spain project.
\end{acknowledgments}
\section*{Data Availability}
The data that support the findings of this article are openly available \cite{Dataset}.
%

\widetext

\begin{center}
\textbf{\large Appendix A: The dimensional reduction}\label{Appendix_A_restoring_gauge_invariance}
\end{center}
\setcounter{section}{0}
\setcounter{equation}{0}
\setcounter{figure}{0}
\setcounter{table}{0}
\makeatletter
\renewcommand{\theequation}{A\arabic{equation}}
\renewcommand{\thefigure}{A\arabic{figure}}
In this Section, we detail some passages for the dimensional reduction procedure. 

The total derivative produced by $\mathcal{L}_{\mathrm{CS}}$ due to a local transformation $A_\mu\to A_\mu+\partial_\mu f$ reads
\begin{equation}
D_\kappa[A_\mu,f]=\frac{\varepsilon^{\mu\nu\rho}}{2\kappa}\partial_\mu\left(f\partial_\nu A_\rho\right)=\frac{\partial_t}{2\kappa}(f\varepsilon^{ij}\partial_iA_j)-\frac{\varepsilon^{ij}}{2\kappa}\partial_i(fE_j),
\end{equation}
with $E_i=\partial_tA_i-\partial_iA^0$ minus the electric field. We can discard the total time derivative because time is not bounded, so one only needs to subtract $D_\alpha[A_\mu,\mathcal{B}]=-\varepsilon^{ij}\partial_i(\mathcal{B}E_j)/2\alpha$ from $\mathcal{L}_{\mathrm{CS}}$ to restore the gauge invariance with the law of transformation $A_\mu\to A_\mu+\partial_\mu f$, $\mathcal{B}\to\mathcal{B}+\alpha f/\kappa$. Then, one can use Stokes theorem to rewrite the integral over the disk of $-D_\alpha$ as a circulation along the ring of $\mathcal{B}E_\phi$, and rename the electric field as a one-dimensional field strength $\varepsilon^{\mu\nu}F_{\mu\nu}$, $\mu=t,\phi$ to recover the BF term as written in \eqref{eq:S_CSBF}.

Next, in order to see the change of boundary conditions for the bulk equations of motion, we first write the extended Lagrangian in the first order formalism, which allows to identify the flux attachment as the local conservation law and the temporal component as a Lagrange multiplier
\begin{equation}
    \mathcal{L}_{\mathrm{CS+BF}}=\frac{A^0}{\kappa}(\varepsilon^{ij}\partial_iA_j)+\frac{\varepsilon^{ij}}{2\kappa}(\partial_tA_i)A_j-\frac{\varepsilon^{ij}}{2\kappa}\partial_i(A^0A_j)+\frac{\varepsilon^{ij}}{2\alpha}\partial_i(\mathcal{B}E_j).\label{eq:CSBF_Firstorder}
\end{equation}
From this, we also identify the boundary Lagrangian
\begin{equation}
    \mathrm{d}\mathcal{L}_{\mathrm{CS+BF}}[\mathcal{B},A_\mu]=\frac{\varepsilon^{ij}}{2\kappa}\partial_i\left(\frac{\kappa}{\alpha}\mathcal{B}E_j-A^0A_j\right),\label{eq:CSBF_D}
\end{equation}
which is the set of total spatial derivatives in \eqref{eq:CSBF_Firstorder}. We analyze the conservation law it entails under the condition of $A^0$ being a Lagrange multiplier, as it is in the bulk. So we rewrite it in the first order formalism
\begin{equation}
    \mathrm{d}\mathcal{L}_{\mathrm{CS+BF}}[\mathcal{B},A_\mu]= \frac{\varepsilon^{ij}}{2\kappa}\partial_i\bigg( -A^0\left(A_j-\frac{\kappa\partial_\phi}{\alpha r_0}\mathcal{B}\right) +\frac{\kappa}{\alpha} \mathcal{B}\partial_tA_j\bigg),
\end{equation}
and read the conservation law as 
\begin{equation}
\kappa\partial_\phi\mathcal{B}(t,\phi;r_0)=\alpha r_0A_\phi(t,\phi;r_0).\label{eq:constraint_boundary}
\end{equation}
This condition for $A_\phi$ cannot be satisfied for arbitrary $\mathcal{B}$, i.e., $\mathcal{B}$ must be compatible with the flux attachment \eqref{eq:flux_attachment_vacuum}. To see this, one rewrites the gauge field as $A_j=\varepsilon_{jk}\partial_k \beta_0$, so that the flux attachment becomes a Poisson equation
\begin{equation}
    \nabla^2 \beta_0=0\label{eq:Poisson_vacuum},
\end{equation} and \eqref{eq:constraint_boundary} reads $\kappa\partial_\phi\mathcal{B}=-\alpha r_0\partial_r \beta_0$, i.e., a Von Neumann boundary condition for $\beta_0$. For the Poisson equation to admit solutions under this last condition, it must be that $\mathcal{B}$ is periodic. Indeed, using the divergence theorem, 
\begin{equation}
    \oint\partial_\phi\mathcal{B}=-\alpha\kappa^{-1}r_0\oint\partial_r \beta_0=-\alpha\kappa^{-1}r_0\int\mathrm{d}\mathbf{x}\nabla^2 \beta_0=0,\label{eq:Poisson_compatibility}
\end{equation}
from which $\mathcal{B}(r_0,2\pi)=\mathcal{B}(r_0,0)$. The boundary conservation law \eqref{eq:constraint_boundary} is fundamental to send the Chern-Simons Lagrangian into the chiral boson action for $\mathcal{B}$. Then, it is not obvious that the flux attachment \eqref{eq:flux_attachment_vacuum} is solved by pure gradients $A_i=\partial_i\beta$ in our case due to the presence of the boundary that changes the shape of the solution in the bulk. Here, we show that this is still the case. Consider Green's second~identity
\begin{equation}
\int_{\mathcal{D}}\mathrm{d}\mathbf{y}\left(\beta_0(\mathbf{y})\nabla^2G(\mathbf{y})-G(\mathbf{y})\nabla^2\beta_0(\mathbf{y})\right)=\oint_{\partial\mathcal{D}}\left(\beta_0\partial_r G-G\partial_r\beta_0\right)
\label{eq:Green2nd},
\end{equation}
which holds for any $\beta_0, G\in\mathrm{C}^2(\mathcal{D})$. When $G$ is the propagator $\nabla^2_\mathbf{y}G(\mathbf{y},\mathbf{x})=\delta(\mathbf{y}-\mathbf{x})$, then this identity represents the general solution for Poisson's equation in a closed manifold, so that if the propagator is known then $\beta_0$ also is. In our problem, $\nabla^2\beta_0$ is given by the rewriting of the flux attachment \eqref{eq:Poisson_vacuum} and $\partial_r\beta_0$ is fixed by \eqref{eq:constraint_boundary}. Instead, there are no boundary fixed conditions for the propagator, so it is convenient to ask that $\partial_rG=0$ to simplify the right hand side of \eqref{eq:Green2nd}. The condition $\partial_rG=0$ is compatible only with a propagator satisfying $\nabla^2_\mathbf{y}G(\mathbf{y},\mathbf{x})=\delta(\mathbf{y}-\mathbf{x})-\vert\mathcal{D}\vert^{-1}$, because the quantity $\nabla^2G$ in $\mathcal{D}$ and $\partial_rG$ in $\partial\mathcal{D}$ are related through the divergence theorem. The propagator for a disk satisfying these conditions is known \cite{DiBenedetto2009}:
\begin{equation}
    G(\mathbf{y}, \mathbf{x}) = \frac{1}{2\pi} 
    \left( \ln |\boldsymbol{\xi} - \mathbf{y}| \frac{|\mathbf{x}|}{r_0} + \ln |\mathbf{x} - \mathbf{y}| \right) 
    + \frac{1}{4\pi r_0^2} |\mathbf{y}|^2.\label{eq:Propagator}
\end{equation}
Here, $\boldsymbol{\xi}=\mathbf{x}r_0^2/\vert\mathbf{x}\vert^2$ is the reflection of $\mathbf{x}$ across the boundary of the disk.
Finally, by proving the existence of a function $\overline{G}$ such that $\partial_{x_1}G=\partial_{x_2}\overline{G}$, $\partial_{x_2}G=-\partial_{x_1}\overline{G}$, one proves that $A_j=\varepsilon_{jk}\partial_k\beta_0$ can be also written as a pure gradient $A_j=-\partial_j\beta_0$ when $\beta_0$ is expressed in terms of the propagator $\overline{G}$. This function is
\begin{equation}
    \overline{G}(\mathbf{y},\mathbf{x})=\frac{1}{2\pi}\left(\mathrm{arg}(\boldsymbol{\xi}-\mathbf{y})+\mathrm{arg}(\mathbf{x}-\mathbf{y})+2\mathrm{tan}^{-1}(\frac{x_2 y_1 - x_1 y_2}{-r_0^2 + x_1 y_1 + x_2 y_2})\right).\label{eq:Gtilde}
\end{equation}
Therefore, the pure gauge solutions of the flux attachment can be written as in the main text \eqref{eq:CS_puregauge_solutions} with $\beta=-\beta_0$.

We now detail the calculations which lead to \eqref{eq:CS_post_int}. We plug the solution \eqref{eq:CS_puregauge_solutions} in the Chern-Simons part of the Lagrangian \eqref{eq:CSBF_Firstorder}. All terms proportional to $A^0$ then vanish and only the second term survives. By means of the Leibniz rule, it can be rewritten as 
\begin{equation}    \mathcal{L}_{\mathrm{CS}}\overset{\mathrm{sol}}{=}\frac{\varepsilon^{ij}}{2\kappa}\partial_i(\partial_t\beta\partial_{j}\beta)-\frac{\varepsilon^{ij}}{2\kappa}\partial_t\beta\partial_i\partial_j\beta. \label{eq:CS_Sform}
\end{equation}
Since the solenoid solution is time independent, all the corresponding terms are either zero or proportional to a total time derivative, therefore they have been dropped. Also, since $\beta$ is a regular function, the second term in \eqref{eq:CS_Sform} vanishes. The spatial integral of the first term over the disk can be rewritten as the chiral boson action on the boundary by virtue of Stokes theorem, thus arriving at \eqref{eq:CS_post_int}.

Now, in the case in which we start with non-relativistic matter within the disk, we can still work out the same procedure, with few changes. First, the flux attachment condition
\begin{equation}
\varepsilon^{ij}\partial_iA_j=\kappa n,
\end{equation}
can still admit solutions under the boundary condition \eqref{eq:boundary_condition}. Following the same logic as in \eqref{eq:Poisson_compatibility}, i.e to read the flux attachment as the Poisson's equation $\nabla^2\beta_0=-\kappa n$, we must then require that $\mathcal{B}$ fulfills 
\begin{equation}
    \oint\partial_\phi\mathcal{B}=\alpha r_0 \int \mathrm{d}\mathbf{x}n(\mathbf{x}).
\end{equation} 
Then, we solve the flux attachment within the bulk, again in the temporal gauge and with the shape $A_i=\partial_i\beta$, where
\begin{equation}
\beta=-\kappa\int\mathrm{d}\mathbf{y}\overline{G}(\mathbf{y},\mathbf{x})n(\mathbf{y}).
\end{equation}
The integration of Chern-Simons Lagrangian along this solution works in the same way, but the second term in \eqref{eq:CS_Sform} does not go to zero, but gives $-n\partial_t\beta$. The full spatial integral of this term can be written as
\begin{equation}
\int_\mathcal{D}\mathrm{d}\mathbf{x}\mathrm{d}\mathbf{y}\overline{G}(\mathbf{y},\mathbf{x})\partial_tn(\mathbf{y})n(\mathbf{x})=\kappa \int^{r_0^-}\mathrm{d}\mathbf{x}\mathrm{d}\mathbf{y}\overline{G}(\mathbf{y},\mathbf{x})\partial_tn(\mathbf{y})n(\mathbf{x})+\frac{\kappa r_0^2}{2\pi}\int\mathrm{d}\varphi_x\mathrm{d}\varphi_y(-\pi+2(\varphi_x-\varphi_y))\partial_tn(\varphi_y;r_0)n(\varphi_x;r_0).
\end{equation}
By splitting the integral, we see that the second term can be rewritten (by discarding total time derivatives) as proportional to the total particle number on the ring. If such a quantity is conserved, then the second term is zero, and this energy belongs to the bulk and does not live on the ring. In order to have the total particle number to be conserved within the disk, a condition must be fixed for the radial current. Indeed, if we consider the integral of the continuity equation $\partial_tn=-\nabla\cdot\mathbf{J}$ over the disk, we see that the change in time of total particle number $\partial_tN$ is related to $\oint_{\partial\mathcal{D}} J_r$ by virtue of the divergence theorem. Therefore, one must set this quantity to zero in order for matter not to leak out of the disk. If one applies the same rationale to the integral of the density over the ring, we can conserve this quantity alone by fixing the boundary condition $\partial_r J_r\vert_{r_0}=0$. Lastly, we treat the matter Lagrangian and reduce it to the ring. By assuming that the field obeys the Schr\"odinger equation $i\partial_t\Psi=-(\nabla-i\mathbf{A})^2\Psi/2m$ within the disk, and given that $A^0=0$, i.e., it can be added freely, then 
\begin{equation}
    S_{M}\to\int \mathrm{d}t\oint \Psi^*(i\partial_t-A^0)\Psi+\frac{1}{2m}\Psi^*(\nabla-i\mathbf{A})^2\Psi,
\end{equation}
i.e., this action is zero when evaluated on its solution.  In
addition to the request that the total number of particles is conserved on the ring, one can set $(\partial_r+iA_r)\Psi=0$ on the ring. This condition is a boundary condition for the matter radial derivatives, thus it gives specifications to which Cauchy problem the matter evolution obeys to. These conditions guarantee that there is no flow of matter from the ring to the disk.  With the last one, the initial two-dimensional Schr\"odinger Lagrangian goes into a one-dimensional one living on the ring, and couples to the chiral BF Lagrangian.

\begin{center}
\textbf{\large Appendix B: Higher-band correction to the density-dependent flux}\label{Appendix_B_perturbative_corrections}
\end{center}
\setcounter{section}{0}
\setcounter{equation}{0}
\setcounter{figure}{0}
\setcounter{table}{0}
\makeatletter
\renewcommand{\theequation}{B\arabic{equation}}
\renewcommand{\thefigure}{B\arabic{figure}}

In the main text, the derivation of the effective chiral BF model is based on a lower-band truncation of the Raman-dressed gas. However, lower- and higher-band states are coupled by nonsymmetric interatomic interactions. Thus, a small fraction of the atoms in the ground state populates upper-band states, leading to a correction of the effective lower-band theory. In this Appendix we quantify such a correction.

In the dressed frame \eqref{eq:Rdress}\eqref{eq:Rmat}, the full interaction Hamiltonian \eqref{eq:contact_interactions_positionspace} reads
\begin{align}
&\hat{H}_{\mathrm{int}}=\frac{1}{2}\int\prod_i\mathrm{d}\mathbf{k}_{i\perp}\sum_{\{m_i\}_1^4}\sum_{n'_{1}n'_{2}n_1n_2=\pm}g_{n'_{1}n'_{2}n_1n_2} \delta(m_4+m_3-m_2-m_1)\hat{\Psi}_{\mathbf{k}_4n'_2}^\dagger\hat{\Psi}_{\mathbf{k}_3n'_1}^\dagger\hat{\Psi}_{\mathbf{k}_2n_2}\hat{\Psi}_{\mathbf{k}_1n_1},\\ & g_{n'_{1}n'_{2}n_1n_2}=\sum_{\sigma_1,\sigma_2}\hat{R}^\dagger_{n'_2\sigma_2}(m_4)\hat{R}^\dagger_{n'_1\sigma_1}(m_3)g_{\sigma_1\sigma_2}\hat{R}_{n_2\sigma_2}(m_2)\hat{R}_{n_1\sigma_1}(m_1).
\label{eq:int_ham_rotated}
\end{align} 
We now assume that the upper band population is very small compared to the lower band one. With this, we neglect the processes that are quadratic or higher in the upper band field operator, and consider only the processes \begin{align}\label{eq:int_ham_rotated_trunc}
\hat{H}_{\mathrm{int}} \approx &\frac{1}{2}\int\prod_i\mathrm{d}\mathbf{k}_{i\perp}\sum_{\{m_i\}_1^4}\delta(m_4+m_3-m_2-m_1) \big\{g_{--}\hat{\Psi}\dgr_-\hat{\Psi}\dgr_-\hat{\Psi}_-\hat{\Psi}_-+g_\gamma^{(1)}\hat{\Psi}\dgr_+\hat{\Psi}\dgr_-\hat{\Psi}_-\hat{\Psi}_-+g_\gamma^{(2)}\hat{\Psi}\dgr_-\hat{\Psi}\dgr_-\hat{\Psi}_-\hat{\Psi}_+\big\},
\end{align}
with $g_{--}$ already taken into account in the lower band Hamiltonian \eqref{eq:Hlab}, and 
\begin{align}
&g_\gamma^{(1)}=2\left(S_1S_3(g_{\downarrow\downarrow}S_2C_4-g_{\downarrow\uparrow}C_2S_4)-C_1C_3(g_{\uparrow\uparrow}C_2S_4-g_{\uparrow\downarrow}S_2C_4)\right),\cr &g_\gamma^{(2)}=2\left(S_2S_4(g_{\downarrow\downarrow}C_1S_3-g_{\downarrow\uparrow}S_1C_3)-C_2C_4(g_{\uparrow\uparrow}S_1C_3-g_{\uparrow\downarrow}C_1S_3)\right),
\end{align}
where the polarization coefficients are defined in \eqref{eq:polarization_coeff}. In order to quantify the effects of the corrections $\propto g_\gamma$ to the lower-band theory, we explicitly use a mean-field ansatz for the ground state, that is, a condenstate that macroscopically occupies the lower-band state at quasi (angular) momentum $Q$. Due to momentum conservation, the processes in \eqref{eq:int_ham_rotated_trunc} can only couple this condensate to upper-band population at the same momentum. Therefore, we can write the mean-field ansatz as a plane wave of the form
\begin{equation}
\left(\hat{\Psi}_+^\dagger,\hat{\Psi}_-^\dagger\right) \rightarrow \enum{i Q \phi}\left(\sqrt{n_-}\enum{i\theta_-} ,\sqrt{n_+}\enum{i\theta_+}\right),
\end{equation}
where $n_\pm$ is the density in the upper and lower band states. We substitute this mean-field ansatz into the many-body Hamiltonian density and shift the zero of the energy to $E_-(Q)$ to obtain
\begin{equation}\label{eq:Emf_shifted}
\mathcal{E}_\mathrm{mf} \approx  \tilde{\Omega}(Q)n_{+} + \frac{1}{2}g_0 n_-^2 + 2\frac{Q}{r_0} n_-\left(A_S  + \frac{\lambda}{2} n_- \right) + 2 g_\gamma(Q)  n_{-}^{3 / 2} \sqrt{n_{+}} \cos(\Delta\theta),
\end{equation}
with $\Delta\theta = \theta_{-}-\theta_{+}$ and 
$g_\gamma(Q)  = g_\gamma ^{(1)} (Q,Q,Q,Q)$. For simplicity, we will assume $n_- \approx n$ is not varying significantly, and leave $n_+$ and $\Delta\theta$ as the only free variables. Then, minimizing the mean-field energy 
yields
\begin{equation}
n_{+} = \frac{g_{\gamma}^2(Q)  n^3}{\tilde\Omega^2(Q)}, \quad \cos{(\Delta\theta)}=-\mathrm{sign}(g_\gamma).
\end{equation}
In this way, the contribution to the mean-field energy coming from upper-band population reads as an attractive three-body interaction
\begin{equation}\label{mf_energy_upperband}
\mathcal{E}_\mathrm{mf}^{+} \approx - n^{3} \frac{ g_{\gamma}^2(Q) }{ \tilde\Omega(Q)},
\end{equation}
Now, given the definition of the flux $\tilde{\omega}$ \eqref{eq:magnetic_flux_variable} and how it contributes to the mean-field energy density with the lower band population \eqref{eq:Emf_shifted} as $\mathcal{E}^{\mathrm{flux}}_{\mathrm{mf}}=2Qn_-\tilde{\omega}/r_0^2$, we can derive the relationship $\tilde{\omega}=(2n_-)^{-1}r_0^2\partial\mathcal{E}^{\mathrm{flux}}_{\mathrm{mf}}/\partial{Q}$, and approximate again $n_-\approx n$. The corrected magnetic flux then reads
\begin{equation}\label{eq:mf_flux_corrected}
\tilde \omega \approx r_0 A_{S} + \frac{n \lambda r_0 }{2} - \frac{n^2 r_0^2} {2} \frac{\partial}{\partial Q}\left(\frac{ g_{\gamma}^2}{\tilde\Omega} \right)\bigg\vert_Q,
\end{equation}
which contains a correction proportional to $n^2$ that stems from the upper-band contribution \eqref{mf_energy_upperband}. Given that $Q$ is integer-valued, the derivative should be replaced with a finite increment. Expression \eqref{eq:mf_flux_corrected} is in agreement with the flux retrieved by the numerical results of the Raman-dressed gas model shown in Section \ref{sec:numerics}.
\newpage
\begin{center}
\textbf{\large Appendix C: Bogoliubov spectrum}\label{Appendix_C_bogoliubov_spectrum}
\end{center}
\setcounter{section}{0}
\setcounter{equation}{0}
\setcounter{figure}{0}
\setcounter{table}{0}
\makeatletter
\renewcommand{\theequation}{C\arabic{equation}}
\renewcommand{\thefigure}{C\arabic{figure}}
In this Appendix, we derive the Bogoliubov spectrum and group velocities of the lower band Hamiltonian \eqref{eq:Hlab}. We start by discarding the energy offsets and the kinetic energy in the transverse directions. The Hamiltonian density evaluated at the ring radius $r_0$ then reads
\begin{align}
\hat{\mathcal{H}}=\hat{\Psi}^\dagger(\phi)\left[-\frac{\partial_\phi^2}{M^*r_0^2}+\frac{\lambda}{2}\hat{j}_\phi+\frac{g_0}{2}\left(\hat{\Psi}^\dagger\hat{\Psi}\right)(\phi) \right]\hat{\Psi}(\phi)-2\mathrm{Re}\left[A_S\hat{\Psi}^\dagger(\phi)\frac{-i\partial_\phi}{M^*r_0}\hat{\Psi}(\phi)\right],\label{eq:H_cBF_op}
\end{align}
with $\hat{\Psi}\equiv \hat{\Psi}_-$ the lower-band Bose field operator, $M^*$ the effective mass, $A_S$ the static gauge potential in \eqref{eq:AS}, $\hat{j}_\phi=(iM^*)^{-1}(\hat{\Psi}^\dagger\partial_\phi\hat{\Psi}/r_0-(\partial_\phi\hat{\Psi}^\dagger/r_0)\hat{\Psi})$ the current operator, $\lambda$ the current coupling and $g_0$ the effective scattering strength.
Before proceeding with the standard approach to elementary excitations, we choose to first change the frame of reference, in order to be at rest with the ground state of the system which is assumed to have an angular momentum $Q$. Thus, if $\hat{U}$ is the unitary that defines the new field operator $\hat{\Psi}'=\hat{U}\hat{\Psi}$ for a ground state at rest, then $\hat{U}=e^{-iQ\hat{\phi}}$ and the Hamiltonian becomes
\begin{equation}
\hat{\mathcal{H}}=\hat{\Psi}^\dagger(\phi)\hat{h}_0\hat{\Psi}(\phi)=\hat{\Psi}'^{\dagger}(\phi)\left(\hat{U}\hat{h}_0 \hat{U}^\dagger\right)\hat{\Psi}'(\phi),
\end{equation}
with $\hat{h}_0$ the operator part in the square brackets of \eqref{eq:H_cBF_op}. The computation leads to
\begin{equation}
    \hat{h}_0'=\hat{U}\hat{h}_0 \hat{U}^\dagger=\hat{h}_0+\frac{1}{M^*}\left(\frac{Q^2}{r_0^2}+2\frac{Q}{r_0}(-i\frac{\partial_\phi}{r_0}-A_S)+\lambda \frac{Q}{r_0}\left(\hat{\Psi}'^{\dagger}\hat{\Psi}'(\phi)\right)\right).
\end{equation}
We now proceed with the standard approach, of which we discuss the main passages. First, we drop the prime notation while keeping the analysis in the rotating frame and consider the mean-field approximation to be valid, thus setting $\hat{\Psi}^{(\prime)}(\phi)=\Psi(\phi)\hat{a}_0$, with $\hat{a}_0$ the operator annihilating particle in the condensate state and $\Psi(\phi)$ the condensate wavefunction. This way, we obtain a non-linear Schr\"odinger equation for $\Psi$. To understand how elementary excitations behave on top of the condensate, we consider the substitution $\Psi\to \psi_{\mathrm{eq}}+\delta\psi$, where $\psi_{\mathrm{eq}}$ is the unperturbed ground state, $\delta\psi$ describes excitations. We obtain a set of Schr\"odinger equations in this rotating frame which are further manipulated by considering the \textit{ansatze} 
\begin{align}
    &\psi_{\mathrm{eq}}=e^{-i\mu t}\sqrt{n},\label{eq:psi_eq}\\ &\delta\psi=e^{-i\mu t}\left(e^{-iE t}u(\phi)-e^{iE t}v^*(\phi)\right),
\end{align}
which assumes a homogeneous ground-state density $n$, and where $\mu$ is the chemical potential and $u$ ($v$) is the amplitude of the positive (negative) frequency $E$ component of the excitation. Hamiltonian \eqref{eq:H_cBF_op} is locally translational invariant (although not Galilean invariant) only if one neglects interactions with respect to the kinetic energy, i.e., in the weakly interacting regime which is of our interest. Together with working in the regime where the effective scattering length is positive, this allows the choice of the ground state ansatz $\psi_{\mathrm{eq}}$ to be a good representative. From this choice, we deduce the chemical potential from the stationary non-linear Schr\"odinger equation under \eqref{eq:psi_eq} as
\begin{equation}
\mu=\frac{Q^2}{M^*r_0^2} +\frac{2\lambda Q n -2A_SQ}{M^*r_0} + g_0n
\end{equation}
The non-linear Schr\"odinger equation for the perturbation can be instead split into two coupled differential equations for the amplitudes $u$ and $v$. This set of equations is invariant under translations, so we consider the solutions to have the form
\begin{equation}
    u(\phi)=u_Me^{iq\phi},\quad v(\phi)=v_Me^{iq\phi},
\end{equation}
modulo some normalization. Here, $q$ labels the (quasi)momentum of the excitation. This choice turns the system of differential equations into a algebraic homogeneous one:
\begin{align}
&\left(\frac{(q+Q)^2}{M^*r_0^2}-2A_S\frac{q+Q}{M^*r_0}+2g_0 n+\frac{\lambda n}{M^*r_0}\left(q+4Q\right)-\mu-E\right)u_M-\left(g_0n+\frac{\lambda n}{M^*r_0}\left(-q+2Q\right)\right)v_M=0,\\ & \left(\frac{(-q+Q)^2}{M^*r_0^2}-2A_S\frac{-q+Q}{M^*r_0}+2g_0 n+\frac{\lambda n}{M^*r_0}\left(-q+4Q\right)-\mu+E\right)v_M-\left(g_0n+\frac{\lambda n}{M^*r_0}\left(q+2Q\right)\right)u_M=0.
\end{align}
The system admits non trivial solutions only if the determinant is zero, a condition from which we can obtain the spectrum \eqref{eq:Bog_E}. Assuming that the perturbations are linear, the group velocities \eqref{eq:Bog_V} are obtained as left and right finite increments of the energy with respect to the momentum of the perturbation, evaluated at zero momentum. To get these velocities back in the original frame, we apply a boost which is equal to the ground state momentum. 
\end{document}